\documentclass[twocolumn,aps,notitlepage,nofootinbib,reprint]{revtex4}
\usepackage[usenames,dvipsnames]{color}
\usepackage{amsmath, latexsym}
\usepackage{hyperref}
\usepackage[normalem]{ulem}
\usepackage{amssymb}
\usepackage{mathrsfs}
\usepackage{lmodern}
\usepackage[percent]{overpic}
\usepackage[T1]{fontenc}
\usepackage[utf8]{inputenc}
\usepackage{graphicx}
\usepackage{booktabs}
\usepackage{ragged2e}
\usepackage{tabularx}

\definecolor{orange}{rgb}{0.9,0.5,0}
\definecolor{maroon}{rgb}{.8,0,0}
\definecolor{purple}{rgb}{0.8,0.4,0.8}
\definecolor{gray}{rgb}{0.8242,0.8242,0.8242}
\definecolor{dodgerblue}{rgb}{0.12, 0.56, 1.0}
\definecolor{forestgreen}{rgb}{0.2, 0.6, 0.2}

\begin{document}

\title{Improved Cauchy-characteristic evolution system for high-precision numerical relativity waveforms}

\author{Jordan Moxon}
\affiliation{Theoretical Astrophysics, Walter Burke Institute for Theoretical Physics, California Institute of Technology, Pasadena, CA 91125, USA.}

\author{Mark A. Scheel}
\affiliation{Theoretical Astrophysics, Walter Burke Institute for Theoretical Physics, California Institute of Technology, Pasadena, CA 91125, USA.}

\author{Saul A. Teukolsky}
\affiliation{Theoretical Astrophysics, Walter Burke Institute for Theoretical Physics, California Institute of Technology, Pasadena, CA 91125, USA.}
\affiliation{Cornell Center for Astrophysics and Planetary Science,
  Cornell University, Ithaca, New York 14853, USA.}

\newcommand{\todo}[1]{\textcolor{orange}{\texttt{TODO: #1}}} 
\providecommand{\JM}[1]{{\textcolor{forestgreen}{{JM: #1}}}}
\providecommand{\MS}[1]{{\textcolor{Cerulean}{{#1}}}}
\providecommand{\ST}[1]{{\textcolor{dodgerblue}{{ST: #1}}}}
\newcommand{\Note}[1]{\textcolor{blue}{\textbf{[#1]}}}
\newcommand{\red}[1]{{\color{red}{#1}}}


\providecommand{\bs}[1]{\mathring{#1}}
\providecommand{\bl}[1]{{#1}}
\providecommand{\ca}[1]{{#1}^\prime{}}
\providecommand{\rn}[1]{\underline{#1}}
\providecommand{\na}[1]{\breve{#1}}
\providecommand{\ifc}[1]{\hat{#1}}

\newcolumntype{Y}{>{\RaggedRight\arraybackslash}X}

\begin{abstract}

  We present several improvements to the
  Cauchy-characteristic evolution procedure
 that generates high-fidelity gravitational waveforms at
  $\mathcal{I}^+$ from numerical relativity simulations.
  Cauchy-characteristic evolution combines an interior solution of the
  Einstein field equations based on Cauchy slices with an exterior solution
  based on null slices that extend to $\mathcal{I}^+$.
 The foundation of our improved algorithm is a comprehensive method of handling
 the gauge transformations between the arbitrarily specified
 coordinates of the interior Cauchy evolution
     and the unique (up to BMS transformations) Bondi-Sachs
  coordinate system of the exterior characteristic evolution.  
  We present a reformulated set of characteristic evolution equations
  better adapted to numerical implementation.
 In addition, we develop a method to ensure that the angular coordinates used
  in the volume during the characteristic evolution are asymptotically inertial.
  This provides a direct route to an expanded set of waveform outputs and
  is guaranteed to avoid pure-gauge logarithmic dependence that has caused
  trouble for
  previous spectral implementations of the characteristic evolution
    equations.
 We construct a set of Weyl scalars compatible with the Bondi-like coordinate
  systems used in characteristic evolution, and determine simple, easily
  implemented forms for the asymptotic Weyl scalars in our suggested set of
  coordinates.
\end{abstract}

\maketitle


\section{Introduction and Motivation} \label{sec:introduction}

In the years since the original observations of the gravitational waves from
compact binary mergers \cite{Abbott:2016blz, Abbott:2016nmj}, sensitivity
improvements of ground-based gravitational wave detectors have steadily
increased the detection rate and the quality of data observed from merger events
\cite{LIGOScientific:2018mvr, Aasi:2013wya}.
The detection and analysis of the resulting gravitational wave data require
precise predictions of the signal from general relativity or beyond-GR theories.
These predictions are used to extract faint gravitational wave signals from data
contaminated by detector noise \cite{Hanna:2010zzb, Flanagan:1997sx} , and to
determine the detailed nature of the binary merger
\cite{Abbott:2016apu,Kumar:2018hml,Kumar:2015tha, Lange:2017wki,
  Lovelace:2016uwp}, including the parameters of the participating compact
objects as well as constraints on parameters that describe deviations from
general relativity \cite{LIGOScientific:2019fpa,Isi:2019aib}.

The three principal techniques for generating gravitational waveform predictions
are post-Newtonian theory, self-force perturbation theory, and numerical
relativity.
Each of these techniques is most effective in a particular region of parameter
space of a compact binary coalescence.
Post-Newtonian theory is accurate and efficient when relativistic effects are
small, so it models a binary at large separation well.
Self-force techniques analytically expand the Einstein field equations in powers
of the mass ratio, so they
work well when describing a binary inspiral with disparate mass
scales.
Finally, numerical relativity is most useful when describing the strong-field
region of comparable or intermediate mass ratio binary coalescence.
In particular, the loudest part of the compact object coalescences observed by
ground-based detectors aLIGO and VIRGO \cite{Hanna:2010zzb} are of the type for
which numerical relativity simulations are most valuable.
As the rate and extracted signal-to-noise ratio of gravitational wave detections
improves, it will become increasingly important to have highly precise waveform
predictions to make best use of rich gravitational waveform data.

Multiple successful numerical relativity code bases have been constructed to
determine the evolving metric of a compact binary coalescence \cite{SpEC,
  Loffler:2011ay, Bruegmann:2003aw, Ruchlin:2017com}.
Numerical relativity simulations operate by evolving Cauchy data on a finite
spatial domain with approximate outgoing radiation boundary conditions.
It is typically necessary to choose a gauge that makes the evolution stable and
efficient, as opposed to choosing a gauge that provides convenient coordinates
for interpreting the results.
A key result of a successful Cauchy simulation is the spacetime metric and its
derivatives on one or more timelike worldtube surfaces at a chosen distance from
the source, often ${}\sim 100-1000$ times the Schwarzschild radius of the
compact companions.

The worldtube data produced by a Cauchy simulation is then used as input to a
separate calculation to determine the gravitational wave observables, such as
the time-dependent strain, asymptotically far from the system.
The simplest method is to use linearized perturbation theory to derive an
approximate asymptotic waveform by assuming that the nonlinear effects of
general relativity are sufficiently small
at the worldtube 
to be neglected.
A more precise method, known as waveform extrapolation, is to take advantage
of the peeling theorem by computing the Weyl scalar $\psi_4$ at a sequence of
concentric worldtubes, fitting to a power series in $r^{-1}$, and using the
$r^{-1}$ piece of $\psi_4$ to compute the asymptotic
waveform \cite{Boyle:2009vi, Chu:2015kft}.

The most faithful method, however, is to take the worldtube data provided by a
Cauchy evolution as the inner boundary data for a second nonlinear evolution of
the Einstein field equations.
This second evolution is based on null (or ``characteristic'') hypersurfaces
rather than spacelike hypersurfaces, and each null hypersurface extends fully to
future null infinity $\mathcal{I}^+$.

This method of obtaining boundary data for the null foliation from an interior
Cauchy evolution and propagating the gravitational wave information via an outer
nonlinear evolution to $\mathcal{I}^+$ is known as \emph{Cauchy-characteristic
  evolution}\footnote{The acronym CCE has also been used in the past to refer to
  ``Cauchy-characteristic extraction'', which describes only the part of the
  computation moving from the Cauchy coordinates to a set of quantities that
  could separately be evolved on null characteristic curves.
 Most of our descriptions refer to the entire algorithm as a single part of the
  wave computation, so we refer to the combination of Cauchy-characteristic
  extraction and characteristic evolution as simply CCE.
} (CCE).
Historically, the development of CCE has proceeded by first developing an
evolution method on null hypersurfaces \cite{Bishop:1996gt, Bishop:1997ik}, then
creating a system for obtaining boundary data for the characteristic evolution
determined by worldtube data from a Cauchy evolution \cite{Winicour:2012znc,
  Babiuc:2011qi, Bishop:1998uk, Reisswig:2009rx, Reisswig:2009us}.
CCE was first implemented as a finite-difference code in Pitt Null
\cite{Babiuc:2010ze, Bishop:1998uk}, for which first results were presented in
\cite{Reisswig:2009rx,Reisswig:2009us}.
More recently, a spectral version of the algorithm was developed as a module in
SpEC \cite{Handmer:2014qha, Handmer:2015dsa, Handmer:2016mls, Barkett:2019uae}.

The CCE system evolves the Einstein field equations on a foliation of null
hypersurfaces, each with an inner boundary on the timelike worldtube provided by
the the Cauchy evolution.
Introducing a compactified radial coordinate system
allows the system to evolve on a finite spherical-shell
  domain with an outer boundary corresponding to $\mathcal{I}^+$.
CCE is particularly powerful in determining subdominant angular modes of the
gravitational wave signal, including slowly-varying ``memory'' modes, which are
hoped to be detectable from aggregation of several merger observations.
It is anticipated that modern advanced LIGO merger detections could reach the
point where a memory signal is detectable with a signal to noise
ratio of $\sim 5$ with $\sim 90$ detections similar to GW150914 \cite{Lasky:2016knh}.

A significant challenge in working with the CCE system is caused by the gauge of
the metric data on the worldtube, which is determined from the Cauchy evolution.
The interior Cauchy evolution is typically forced to choose its coordinates to
be convenient for timelike evolution in the strong-field region, and is under no
constraint to provide the worldtube data in a gauge for which the coordinate
representation of the metric falls off to a standard Minkowski form at
$\mathcal{I}^+$.
In practical implementations of the CCE system, it is frequently found that in
the Cauchy evolution's gauge, all components of the metric that are not
completely fixed by additional coordinate transformations possess nontrivial
values at $\mathcal{I}^+$.

In the canonical treatment of the CCE system, the metric is put into a form that
resembles the Bondi-Sachs form of the metric (see
Eq.~(\ref{eq:BondiSachsMetric}) below), but for which each of the metric
components is constrained only to have finite value at $\mathcal{I}^+$
rather than the precise falloff behavior (see Eq.~(\ref{eq:BondiSachsFalloffs})
below) required by Bondi-Sachs coordinates.
For the remainder of this paper, we will refer to the canonical CCE form of the
metric as ``Bondi-like'' to distinguish those coordinates from a true
Bondi-Sachs gauge.
Our notation choices for the different forms of the metric are detailed in
Section \ref{sec:foundations}.

The first implementation of CCE, PittNull \cite{Bishop:1997ik}, successfully
evolved the characteristic system on the compactified asymptotic domain, but as
a finite-difference implementation it was not efficient enough to be adopted ubiquitously
\cite{Barkett:2019uae}.
The SpEC CCE implementation uses pseudospectral techniques to evolve the
characteristic system, so it obtains precise waveforms far faster
\cite{Handmer:2016mls, Handmer:2014qha, Barkett:2019uae}.
However, the precision of the SpEC spectral method is threatened by the
development of pure-gauge logarithmic terms that arise in CCE implementations
using Bondi-like coordinates.
Such logarithmic dependence disrupts the exponential convergence with resolution
otherwise enjoyed by spectral methods.
Further, previous implementations of CCE have had the capability to extract only
the asymptotic news and not other asymptotic quantities such as the Weyl
scalars, because the expressions for those quantities are extremely complicated
in general Bondi-like coordinates.
Initial explorations with the SpEC CCE implementation \cite{Handmer:2016mls}
  have demonstrated the feasibility of extracting the Weyl scalar $\psi_4$ from
  BMS flux quantities.

In this paper, we introduce a novel strategy for CCE that alleviates the most
prominent remaining challenges for spectral implementations.
Our new method is constructed from a well-chosen gauge restriction placed on the
Bondi-like coordinate system used to perform the evolution.
Under our gauge restriction, the evolution system is provably free of pure-gauge
logarithms and all quantities needed to fix the gauge are easily calculated from
the supplied Cauchy metric on the worldtube.
Further, our gauge restriction alleviates most of the complexity in evaluating
the asymptotic waveform quantities, leading to simpler computation of the
Bondi-Sachs news as well as succinct, practical formulas for the strain and for
the leading contribution to all five Weyl scalars near $\mathcal{I}^+$.

The combination of mathematical results presented in this paper is best
described by their role in the suggested algorithm for the next generation of
CCE implementation:

\begin{enumerate}
\item Initialize the characteristic system using well-chosen data on the initial
  null hypersurface that is compatible with a fully regular evolution procedure
  (thoroughly demonstrated in detail for the first time in Section
  \ref{sec:regularity_preservation})
\item At the intersection of the worldtube and
  the initial null hypersurface, compute  the Bondi-like metric from the supplied Cauchy metric
  on the worldtube. This uses
  methods developed in \cite{Bishop:1998uk} and reviewed in
  Section \ref{sec:foundations}.
\item At the intersection of the worldtube and the initial null
  hypersurface, compute the gauge transformation for a subset of the Bondi-Sachs
  spin-weighted scalars to a new ``partially flat'' gauge presented in this
  work (Section \ref{sec:bondi_transforms}).
 The partially flat gauge is easy to compute from the Bondi-like metric and
  provably avoids the pure-gauge logarithmic dependence (demonstrated in Section
  \ref{sec:regularity_preservation}) that arises in a general Bondi-like gauge.
\item Perform the radial integration from the worldtube to $\mathcal{I}^+$ on
  the initial null hypersurface, for the first subset of Bondi-Sachs scalars
  determined in the previous substep.
 This determines these Bondi-Sachs scalars everywhere on the 
      initial null hypersurface.
 The radial differential equations for this step are improved in Section
  \ref{sec:compactified_evolution} via a coordinate description optimized for
  practical implementation in a spectral domain.
 These equations are equivalent to a coordinate transformation and
  simplification applied to the original derivations \cite{Bishop:1997ik}.
\item At the intersection of the initial null hypersurface
  and $\mathcal{I}^+$, use the results of the previous step to
  obtain information  necessary
  to complete the gauge transformation to partially flat coordinates for the
  remaining Bondi-Sachs spin-weighted scalars
 that were not calculated in steps 3 and 4.
  (see the derivation of the volume coordinates in
  Section \ref{sec:bondi_transforms} and the newly developed procedure in
  Section  \ref{sec:regularity_preservation}).
\item At the intersection of the worldtube and the initial null hypersurface,
  complete the transformation to partially flat coordinates for the final
  collection of Bondi-Sachs spin-weighted scalars (Section
  \ref{sec:bondi_transforms}), and perform the radial integration to obtain the
  remaining Bondi-Sachs spin-weighted scalars everywhere on the initial
  null hypersurface.
\item Perform an ODE integration in partially flat retarded time $u$ to
  obtain data on the next null hypersurface.
  This uses the Bondi-Sachs spin-weighted quantities obtained by steps 2--6.
\item  Repeat steps 2--7 for each successive null hypersurface.
\item Use the simple expressions of the Weyl scalars in our new gauge (Section
  \ref{sec:weyl_scalars}) to produce detailed gauge-invariant (up to BMS
  freedom) information about the dynamic spacetime from the spin-weighted
  scalars evaluated in the evolution procedure.
\end{enumerate}
The full detailed implementation strategy, complete with references to each of
the equations for the computation steps, can be found in Subsection
\ref{sec:computation_roadmap}.

The numerical implementation of these techniques into an efficient, robust CCE
code will be presented in forthcoming work \cite{CceNumeric}.
The numerical implementation is built into the versatile code base SpECTRE
\cite{Kidder:2016hev}, which targets scalable astrophysical simulation of
various multiphysics systems, including multimessenger compact object
coalescence.
The efficient CCE code we are developing will be an important ingredient in the
production of the gravitational wave predictions from astrophysical events
simulated in SpECTRE.

\section{Foundations of CCE}  \label{sec:foundations}

In this section, we review the current standard methods for CCE implementations.
Thus, this section is entirely a recapitulation of calculations presented in past
  works \cite{Bondi:1962px,Newman:1961qr, Bishop:1997ik, Bishop:1998uk,
    Handmer:2014qha}, adjusted to a notation compatible with the new
  calculations in the remaining sections of this paper.
The CCE algorithm takes as input the spacetime metric and its derivatives on a
chosen 2+1-dimensional surface outside the strong-field region of the Cauchy
simulation.
The CCE procedure then adapts that metric to a Bondi-like form in which the
Einstein field equations are amenable to a hierarchical evolution procedure.
The metric components expressed as Bondi-like spin-weighted scalars are
determined on each null hypersurface that intersects the worldtube at constant
Cauchy simulation time.
A series of integrations along the null hypersurfaces generate the necessary
data for performing a time step of the hyperbolic part of the characteristic
system.
The result of the characteristic evolution then specifies the full metric  at
$\mathcal{I}^+$, which is the outer boundary point of the compactified domain.
Finally, that metric 
can be used in the evolution of inertial coordinates
on $\mathcal{I}^+$ that are used to produce meaningful gravitational wave
observables in a BMS frame selected by the metric on the
initial CCE hypersurface.

In the discussions of this section and throughout the CCE formalism, we find
ourselves with an inconvenient surfeit of coordinate systems.
To assist the reader in making sense of the various coordinates and indices, we
include in the Appendix \ref{app:indices} a table of the coordinates, their
associated index adornment, and where they are used in the paper.
All coordinate systems are considered spherical (possessing one timelike
coordinate, one radial coordinate, and two coordinates on the sphere), Greek
letters are used to indicate spacetime 4-indices, Latin letters in the range
$i\dots n$ are used to indicate spatial 3-indices, and capital Latin letters are
used to indicate angular 2-indices.
Where necessary to clarify the
coordinate dependence of 
spacetime fields, the symbols
will also be decorated by the same adornment used for indices, like $\bs \beta$.

\subsection{Bondi-Sachs metric and Bondi-like coordinates for CCE} \label{sec:bondi_sachs}

In Bondi-Sachs coordinates, an asymptotically flat spacetime possesses a metric
in spherical coordinates $\{\bs u, \bs r, \bs x^{\bs A}\}$ of the form \cite{Bondi:1962px,
  Barnich:2009se, Flanagan:2015pxa},
\begin{align} \label{eq:BondiSachsMetric}
  ds^2 =
  & -\left(e^{2 \bs \beta} \frac{\bs V}{\bs r} - \bs r^2 \bs h_{\bs A  \bs B} \bs U^{\bs A} \bs U^{\bs B} \right) d\bs u^2 
    - 2 e^{2 \bs \beta} d\bs u d\bs r \notag\\
  &- 2 \bs r^2 \bs h_{\bs A \bs B} \bs U^{\bs B} d \bs u d\bs x^{\bs A}
    +\bs r^2 \bs h_{\bs A \bs B} d \bs x^{\bs A} d\bs x^{\bs B}.
\end{align}
In these expressions, $\bs x^{\bs A}$ denotes the pair of angular coordinates.
The Bondi-Sachs coordinates impose the gauge conditions $g_{\bs r \bs r} = 0$,
$g_{\bs r \bs A} = 0$, and the determinant of the angular components is set to the
determinant of the unit sphere metric $q_{\bs A \bs B}$,
\begin{equation}
  \det(h_{\bs A \bs B}) = \det(q_{\bs A \bs B}).
\end{equation}
Further, the metric in Bondi-Sachs coordinates asymptotically approaches the
Minkowski metric.
The asymptotic restriction demands falloff rates for each of the metric
components in (\ref{eq:BondiSachsMetric}) \cite{Bondi:1962px, Barnich:2009se,
  Flanagan:2015pxa}:
\begin{subequations} \label{eq:BondiSachsFalloffs}
  \begin{align}
    \lim_{\bs r\rightarrow\infty}    \bs \beta(\bs x^{\bs \alpha}) &= \mathcal{O}(\bs r^{-1}),\\
    \lim_{\bs r\rightarrow\infty}   \bs V(\bs x^{\bs \alpha}) &= \bs r  + \mathcal{O}(\bs r^0),\\
    \lim_{\bs r\rightarrow\infty}   \bs U^{\bs A}(\bs x^{\bs \alpha}) &= \mathcal{O}(\bs r^{-2}),\\
    \lim_{\bs r\rightarrow\infty}   \bs h_{\bs A \bs B}(\bs x^{\bs \alpha}) &= q_{\bs A \bs B}(\bs x^{\bs A}) + \mathcal{O}(\bs r^{-1}),
  \end{align}
\end{subequations}

The Bondi-Sachs coordinate system is rather restrictive, and transforming to a
Bondi-Sachs coordinate system from a generic metric depends on detailed
information about the metric all the way out to $\mathcal{I}^+$ (see Section
\ref{sec:bondi_transforms} for a concrete description of how to obtain that
information).
For expedience, the standard methods for numerical implementations of CCE do not
place the metric in a true Bondi-Sachs coordinate system, instead settling for
the set of gauge conditions that can be evaluated locally in the bulk of the
spacetime.
The Bondi-like coordinate system used in previous CCE treatments takes the same
form as Eq.~\eqref{eq:BondiSachsMetric},
\begin{align} \label{eq:BondiLikeMetric}
  ds^2 =
  & -\left(e^{2 \bl \beta} \frac{\bl V}{\bl r} - \bl r^2 \bl h_{\bl A  \bl B} \bl U^{\bl A} \bl U^{\bl B} \right) d\bl u^2 
    - 2 e^{2 \bl \beta} d\bl u d\bl r \notag\\
  &- 2 \bl r^2 \bl h_{\bl A \bl B} \bl U^{\bl B} d \bl u d\bl x^{\bl A}
    +\bl r^2 \bl h_{\bl A \bl B} d \bl x^{\bl A} d\bl x^{\bl B}.
\end{align}
The gauge conditions imposed for (\ref{eq:BondiLikeMetric}) are the same local
choices, $g_{\bl r \bl r} = 0$, $g_{\bl r \bl A} = 0$, and
$\det(h_{\bl A \bl B}) = \det(q_{\bl A \bl B})$.
However, the asymptotic dependence is relaxed to the simple requirement that all
metric components are asymptotically finite,
\begin{subequations} \label{eq:BondiLikeFalloff}
  \begin{align}
    \lim_{\bl{r}\rightarrow\infty} \bl \beta(\bl{x}^{\bl{\alpha}})
    = \mathcal{O}(r^0),\\
    \lim_{\bl{r}\rightarrow\infty} \bl W(\bl{x}^{\bl{\alpha}})
    = \mathcal{O}(r^0),\\
    \lim_{\bl{r}\rightarrow\infty} \bl U^{\bl{A}}(\bl{x}^{\bl{\alpha}})
    = \mathcal{O}(r^0),\\
    \lim_{\bl{r}\rightarrow\infty} \bl h_{\bl{A} \bl{B}}(\bl{x}^{\bl{\alpha}})
    = \mathcal{O}(r^0).
  \end{align}
\end{subequations}
As an implementation note, the standard CCE algorithm doesn't even truly
impose (\ref{eq:BondiLikeFalloff}).
Instead, the gauge fixed by most Cauchy worldtube data is sufficiently
well behaved to avoid pathological divergent dependence on radius.
Notably, requirements on Bondi-like coordinate systems are strictly weaker than
the Bondi-Sachs coordinates, so any generic conclusions obtained about a
Bondi-like metric can be immediately applied to a Bondi-Sachs metric.

To ease computations of the angular components of the metric, we introduce the
complex dyad $q^A$ associated with the unit sphere metric $q_{A B}$.
The following techniques may be applied in any coordinate system that satisfies
the determinant condition $\det(h_{A B}) = \det(q_{A B})$, so they
are applicable to
both Bondi-Sachs and Bondi-like coordinate systems.
Here we use unadorned indices for notational simplicity.
First,
\begin{equation}
  q_{A B} = \frac{1}{2} (q_{A} \bar{q}_{B} + \bar{q}_{A} q_{B}).
\end{equation}
Note that under this definition, the dyad has normalization $q^A \bar{q}_A = 2$,
which is chosen for the numerical convenience of avoiding factors of $\sqrt{2}$
in dyad components and derivatives.
For computations in which the angular components must be expanded explicitly,
this paper will make use of standard spherical coordinate angles
$\{\theta, \phi\}$, and select the complex dyad,
\begin{equation} \label{eq:ComplexDyadChoice}
  q^A = \left\{-1, \frac{-i}{\sin\theta}\right\}.
\end{equation}

Each angular component of the Bondi-like metric is contracted with either the
complex dyad $q^A$ or its conjugate $\bar{q}^A$ to form \emph{spin-weighted
  scalar} components \cite{Newman:1961qr, Bishop:1997ik}.
The spin-weight of the scalars is determined by how they transform according to
rotations in the choice of complex dyad $q^A$.
The unit sphere metric is symmetric under in-plane rotations of the complex
dyad
\begin{equation} \label{eq:DyadRotation}
  q^A\rightarrow q^A e^{i \psi}.
\end{equation}
We then refer to a quantity $v$ as possessing a spin-weight $s$ if it
transforms as
\begin{equation}
  v \rightarrow v e^{i s \psi}
\end{equation}
under the dyad rotation (\ref{eq:DyadRotation}).
Consequently, for any angular vector $v^A$, the scalar $v = v^A q_A$ is of
spin-weight 1, and $\bar{v} = v^A \bar{q}_A$ is of spin-weight -1.
We adopt the following standard notation for spin-weighted scalars derived from
angular Bondi-like metric components:
\begin{subequations}
  \begin{align}
    U &\equiv U^A q_A, &\text{(spin-weight 1)}&\\
    Q &\equiv r^2 e^{-2 \beta} q^A h_{A B} \partial_r U^B,  &\text{(spin-weight 1)}& \label{eq:QDef}\\
    r^2 W &\equiv V - r, &\text{(spin-weight 0)}& \label{WDef}\\
    J &\equiv \frac{1}{2} q^A q^B h_{A B},  &\text{(spin-weight 2)}&\\
    K &\equiv \frac{1}{2} q^A \bar{q}^B h_{A B} \notag\\
      &= \sqrt{1 - J \bar{J}},   &\text{(spin-weight 0)}&\label{eq:KDef}
  \end{align}
\end{subequations}
where $Q$ in (\ref{eq:QDef}) is introduced to reduce the Einstein field
equations for the spin-weighted components to only first derivatives in $r$.
The equality in (\ref{eq:KDef}) arises from the normalization of the angular
metric determinant $\text{det}(h_{A B}) = \text{det}(q_{AB})$ from the
Bondi-like construction.

We introduce the spin-weighted angular differential operators $\eth$ and
$\bar{\eth}$.
For any spin-weighted scalar quantity
$v = q_1^{A_1} \dots q_n^{A_n} v_{A_1 \dots A_n}$, where each $q_i$ may be
either $q$ or $\bar{q}$, we define the spin-weighted derivatives,
\begin{subequations} \label{eq:DefEthEthbar}
  \begin{align}
    \eth v = q_1^{A_1} \dots q_n^{A_n} q^B D_B v_{A_1 \dots A_n},\\
    \bar{\eth} v = q_1^{A_1} \dots q_n^{A_n} \bar{q}^B D_B v_{A_1 \dots A_n}.
  \end{align}
\end{subequations}
In (\ref{eq:DefEthEthbar}), $D_A$ denotes the covariant derivative associated
with the unit sphere metric $q_{A B}$.
In particular, for our choice of the complex dyad (\ref{eq:ComplexDyadChoice}),
the spin-weighted derivative operator applied to a scalar $v$ of spin-weight
$s$ takes the coordinate form
\begin{equation}
  \eth v = -(\sin \theta)^s \left(\frac{\partial}{\partial \theta}
    + \frac{i}{\sin \theta} \frac{\partial}{\partial \phi}\right)
  \left[(\sin \theta )^{-s} v\right].
\end{equation}

\subsection{Boundary transformations} \label{sec:boundary_transforms}

The first task of a numerical implementation of CCE is to transform the metric
provided by a Cauchy simulation on a particular worldtube to the Bondi-like form
(\ref{eq:BondiLikeMetric}).
In this section, we review the practical steps that can be used to impose the
gauge conditions $g_{\bl{r} \bl{r}} = 0$, $g_{\bl{r} \bl{A}} = 0$,
$\det(g_{\bl{A} \bl{B}}) = \det(q_{\bl{A} \bl{B}})$ on the input worldtube
$\Gamma$.
The material we review here is similar to the procedures developed in
\cite{Bishop:1998uk}, though we describe the streamlined procedure
\cite{Barkett:2019uae} that does not first extrapolate to a surface of constant
Bondi-like $\bl{r}$.
Instead, the initial data in our description is given at a surface described by
a worldtube radius function $\bl{r} = R(\bl{x}^{\bl{A}})$.

The input to the CCE algorithm is a set of metric components
$g_{\ca \alpha \ca \beta}$ and their first partial derivatives
$g_{\ca \alpha \ca \beta, \ca \gamma}$ on a surface $S_{\ca r}$ of constant
$\ca r$ and $\ca t$ from the Cauchy evolution.
On that surface, we perform the ADM decomposition of the metric components
natural for the Cauchy evolution:
\begin{align}
  ds^2 =& \left(-\alpha^2  + \beta^{\ca i} \beta^{\ca j}
    g_{\ca i \ca j} \right)d\ca t^2 \notag\\
  &+ 2\beta^{\ca i} g_{\ca i \ca j}  d\ca x^{\ca j} d\ca t
  + g_{\ca i \ca j} d\ca x^{\ca i} d\ca x^{\ca j}.
\end{align}
Here $\beta^{\ca i}$ is the shift vector, which is not to be confused with the
metric component $\beta$ in the Bondi-like metric~(\ref{eq:BondiLikeMetric}).

First, we wish to construct a null vector $l^{\ca \alpha}$ to act as the
generators for the outgoing null cone at the worldtube surface.
The normal vector $s^{\ca \alpha}$ to the provided surface $S_{\ca r}$ is
used for the radial component for a candidate null vector.
\begin{subequations}
  \begin{align}
    s^{\ca t} &= 0, \\
    s^{\ca i} &= \frac{g^{\ca i \ca j} \partial_{\ca i}
                   \ca r}{\sqrt{g^{\ca i \ca j}
                   \partial_{\ca i} \ca r
                   \partial_{\ca j} \ca r}}.
  \end{align}
\end{subequations}
Define also the hypersurface normal associated with the Cauchy evolution $n^{\ca \alpha}$
\begin{subequations}
  \begin{align}
    n^{\ca t} &= \frac{1}{\alpha},\\
    n^{\ca i} &= \frac{-\beta^{\ca i}}{\alpha}.
  \end{align}
\end{subequations}
Then, by construction, $n^{\ca \alpha} n_{\ca \alpha} = -1$,
$s^{\ca \alpha} s_{\ca \alpha} = 1$, and
$n^{\ca \alpha} s_{\ca \alpha} = 0$.
Therefore, we determine a normalized null vector $l^{\alpha^\prime}$ as
\cite{Bishop:1998uk}
\begin{equation}
  l^{\alpha^\prime} = \frac{n^{\alpha^\prime}
    + s^{\alpha^\prime}}{\alpha - g_{i^\prime j^\prime} \beta^{i^\prime} s^{j^\prime}}.
\end{equation}
We now construct a new set of null-radius coordinates
$\{\rn u, \rn \lambda, \rn x{}^{\rn A}\}$, where
$\rn \lambda$ is an affine parameter along the null rays generated by
$l^{\alpha^\prime}$.
In the set of null-radius coordinates, we use the time and angular coordinates
from the Cauchy data unchanged 
$\rn x^{\rn A} = \delta^{\rn A}{}_{\ca A}
x^\prime{}^{A^\prime}$,
$\rn u = \ca t$.

The metric components in the null-radius coordinates are
\begin{subequations} \label{eq:NullRadiusMetric}
  \begin{align}
    g_{\rn{\lambda} \rn{u}}
    &= l^{\ca \alpha} g_{\ca \alpha \ca t} = -1,\\
    g_{\rn{\lambda} \rn{\lambda}}
    &= l^{\ca \alpha} l^{\ca \beta} g_{\ca \alpha \ca \beta} = 0,\\
    g_{\rn{\lambda} \rn{A}} 
    &= l^{\ca \alpha} \frac{\partial \ca x^{ \ca i}}{\partial \rn{x}^{\rn{A}}}
      g_{\ca \alpha \ca i}
      = l^{\ca \alpha} \delta_{\rn{A}}{}^{\ca A}
      \frac{\partial x^{\ca i}}{\partial \ca x{}^{\ca A}} g_{\ca \alpha \ca i} = 0,\\
    g_{\rn{u} \rn{u}}
    &= g_{\ca t  \ca t} ,\\
    g_{\rn{u} \rn{A}}
    &= \frac{\partial \ca x^{\ca i}}{\partial \rn{x}^{\rn{A}}} g_{\ca t \ca i}
      = \delta_{\rn{A}}{}^{\ca A}
      \frac{\partial \ca x^{\ca i}}{\partial \ca x{}^{\ca A}} g_{\ca t \ca i},\\
    g_{\rn{A} \rn{B}}
    &= \frac{\partial \ca x^{\ca i}}{\partial \rn{x}^{\rn{A}}}
      \frac{\partial \ca x^{\ca j}}{\partial \rn{x}^{\rn{B}}} g_{\ca i \ca j} \notag\\
      &= \delta_{\rn{A}}{}^{\ca A}  \delta_{\rn{B}}{}^{\ca B}
      \frac{\partial x^{\ca i}}{\partial \ca x{}^{\ca A}}
      \frac{\partial x^{\ca i}}{\partial \ca x{}^{\ca B}} g_{\ca i \ca j}.
  \end{align}
\end{subequations}
At this point, we've determined a suitable null coordinate system for the
worldtube metric, enforcing
$g_{\rn{\lambda} \rn{\lambda}} = g_{\rn{\lambda} \rn{A}} = 0$.
To complete the transformation to coordinates compatible with the Bondi-like
metric form (\ref{eq:BondiLikeMetric}), we must construct an areal radial
coordinate $\bl r$ such that $g_{\bl{A} \bl{B}} = \bl{r}^2 h_{\bl{A} \bl{B}}$
and $\det(h_{\bl{A}\bl{B}}) = \det(q_{\bl{A}\bl{B}})$.

Define the Bondi-like radius,
\begin{equation}\label{eq:BondiLikeRadius}
  \bl{r} = \left[\frac{\det(g_{\rn{A} \rn{B}})}
    {\det(q_{\rn{A} \rn{B}})}\right]^{1/4},
\end{equation}
where, as in the previous subsection, $q_{\rn A \rn B}$ represents the angular metric on
the unit sphere.
The Bondi-like coordinates adopted by the standard Characteristic Extraction
algorithm are $\{\bl{r}, \bl{u}, \bl{x}^{\bl{A}}\}$, where the time and
angular coordinates are again unchanged from the original set determined by the
Cauchy evolution $\bl{u} = \rn{u}$,
$\bl{x}^{\bl{A}} = \delta^{\bl{A}}{}_{\rn{A}}\rn{x}^{\rn{A}}$.

Finally, the form of the Bondi-like metric (\ref{eq:BondiLikeMetric}) may be
used to determine the spin-weighted scalars associated with the transformed
metric.
The Bondi-like scalars are most conveniently determined from the up-index
components of the metric \cite{Bishop:1998uk, Barkett:2019uae}
\begin{subequations} \label{eq:BondiLikeScalarsFromBondiLikeMetric}
  \begin{align}
    \beta &= -\frac{1}{2} \ln(- g^{\bl{u} \bl{r}})
            = - \frac{1}{2} \ln(\partial_{\rn{\lambda}} r),\\
    U &= \frac{ g^{\bl{u} \bl{A}}}{g^{\bl{u} \bl{r}}} q_{\bl{A}},\\
    W &= \frac{1}{r}\left(1 - \frac{g^{r r}}{g^{\bl u \bl r}}\right) \\
    J &= - \frac{1}{2} r^2 q_{\bl{A}} q_{\bl{B}} g^{\bl{A} \bl{B}},\\
    K &= \sqrt{1 + J\bar{J}},\\
    Q &= \bl{r}^2 (J \partial_{\rn{\lambda}} \bar{U}
        + K \partial_{\rn{\lambda}} U).
  \end{align}
\end{subequations}
Here the up-index components of the metric $g^{\bl{\mu} \bl{\nu}}$ can be
determined from Eqs.~(\ref{eq:NullRadiusMetric})
and the coordinate transformation from
$x^{\rn{\mu}}$ to $x^{\bl{\mu}}$; for details see \cite{Barkett:2019uae}.
Note that several terms in Eqs.~(\ref{eq:BondiLikeScalarsFromBondiLikeMetric})
seem to depend on derivatives of
the Jacobian $\partial x^{\bl{\mu}}/\partial x^{\rn{\nu}}$, and so would have
the danger of requiring multiple derivatives of the input metric.
In fact, the characteristic extraction algorithm depends only on first
derivatives of the input metric.
To make the ability to represent all requisite quantities in terms of first
derivatives explicit, note the identities from \cite{Barkett:2019uae}
\begin{subequations}
  \begin{align}
    U_{,\rn{\lambda}}
    =& - \left(g^{\rn{\lambda} \rn{A}}{}_{,\rn{\lambda}}
      + \frac{\bl{r}_{,\rn{\lambda} \rn{B}}}{\bl{r}_{,\rn{\lambda}}}
      g^{\rn{A} \rn{B}}
      + \frac{\bl{r}_{,\rn{B}}}{\bl{r}_{\rn{\lambda}}} g^{\rn{A} \rn{B}} \right)\notag\\
      &+ 2 \beta_{,\rn{\lambda}} \left(U
      + g^{\rn{\lambda} \rn{A}} q_{\rn{A}}\right),\\
    J_{,\rn{\lambda}} =& -\frac{1}{2} \bl{r}^2 q_{\rn A} q_{\rn B} h^{\rn{A}
                \rn{B}}{}_{,\rn{\lambda}} - \frac{2 \bl{r}_{,\rn{\lambda}}}{r} J.
  \end{align}
\end{subequations}
The term $\beta_{,\rn{\lambda}}$ (which otherwise would depend on the second
derivative $r_{,\rn{\lambda} \rn{\lambda}}$) can be written in terms of only
first derivatives of the input metric by using one of the components of the
Einstein field equations:
\begin{equation}
  \beta_{,\rn{\lambda}} = \frac{\bl{r}}{8 \bl{r}_{,\rn{\lambda}}}
  \left(J_{,\rn{\lambda}} \bar{J}_{,\rn{\lambda}} +
    \left(K_{,\rn{\lambda}}\right)^2\right).
\end{equation}
Finally, we define the helpful quantity $H \equiv \partial_{\bl{u}} J$ on the
worldtube:
\begin{equation}
  H \equiv J_{,\bl{u}} = -\frac{1}{2} \bl{r}^2 q_{\rn A} q_{\rn B} h^{\rn{A}
      \rn{B}}{}_{,\rn{u}} - \frac{2 \bl{r}_{,\rn{u}}}{r} J.
\end{equation}

We emphasize that the set of local computations summarized in this section is
sufficient to enforce the Bondi-like gauge conditions $g_{\bl r \bl r} = g_{\bl r  \bl A} = 0$
and $\det{g_{\bl A \bl B}} = \det{q_{\bl A \bl B}}$, but enforces no restriction on the
asymptotic dependence of the metric components.
In general, the metric in these Bondi-like coordinates will not asymptotically
approach a Minkowski metric, and therefore quantities determined at
$\mathcal{I}^+$ will be in a non-inertial gauge and require an additional
correction described below in Subsection \ref{sec:FoundationsScri}.

\subsection{Hierarchical evolution system} \label{sec:EquationHierarchy}

\begin{figure}[t]
  \includegraphics[width=.5\textwidth]{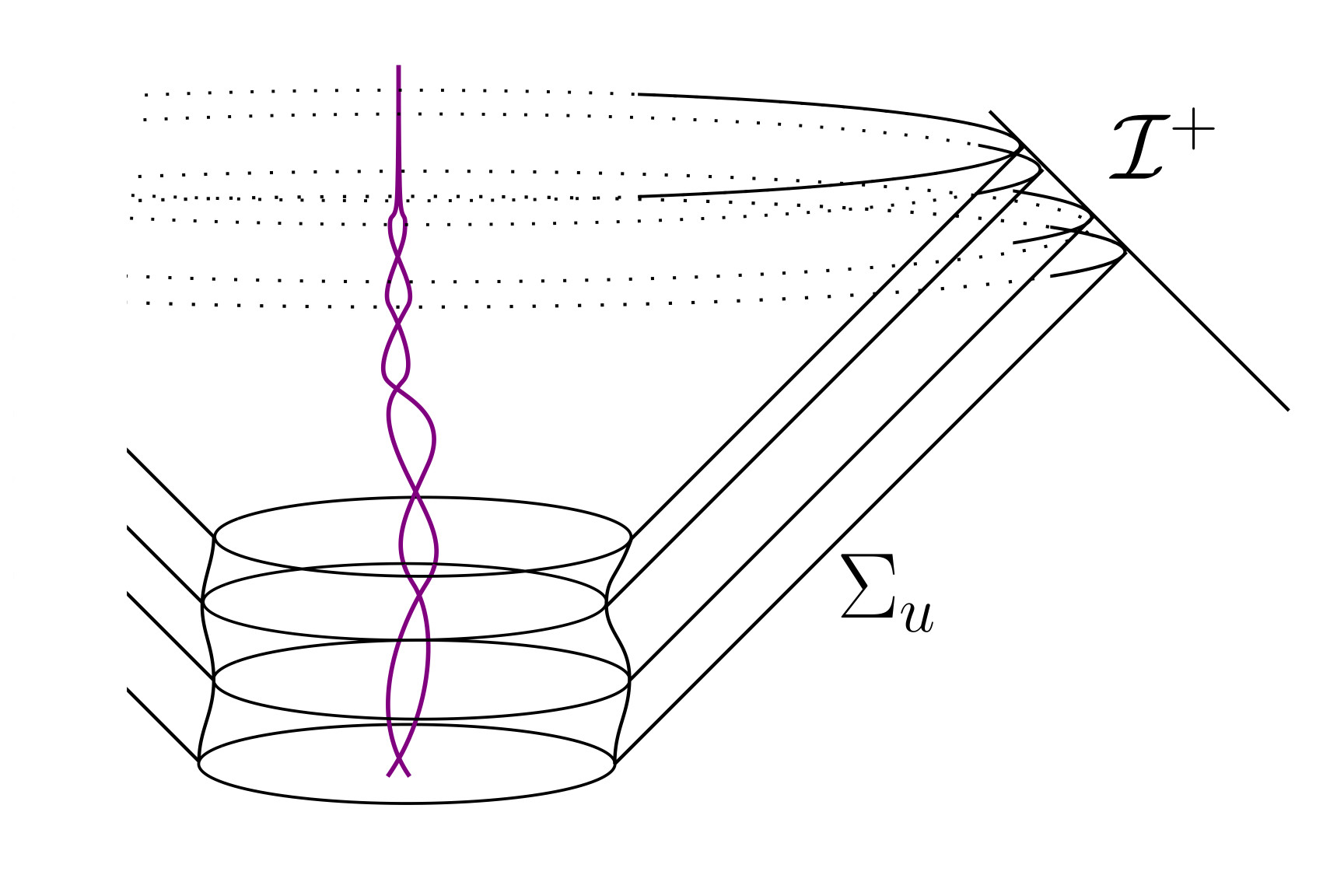}

  \caption{
    An illustration of the Cauchy-characteristic extraction and characteristic
    evolution (CCE) computational domain.
    The central circular region represents the finite spherical domain of the
    Cauchy code.
    CCE is evaluated on the null surfaces extending from the input worldtube to
    $\mathcal{I}^+$.
    Numerical implementations take advantage of a compactified radial coordinate
    to solve for the asymptotic field in a finite numerical domain.
    The characteristic hypersurface equations, Eqs.~(\ref{eq:BondiHierarchy}),
    determine the field values along each $\Sigma_{u}$ using boundary values on
    the worldtube.
    The evolution equation determines the remaining degrees of freedom on
    subsequent hypersurfaces.}\label{fig:HypersurfaceSketch}
\end{figure}

The characteristic evolution proceeds in ascending retarded time $\bar{u}$,
computing the Bondi-like metric on $\bar{u}=\text{constant}$ null hypersurfaces
(see fig.~\ref{fig:HypersurfaceSketch}).
The inputs to this evolution system are the value of $J$ on the initial
hypersurface and the spin-weighted scalars $\beta, U, Q, W$, and $H$ on a
worldtube $\Gamma$ , as determined by the computation reviewed in Subsection
\ref{sec:boundary_transforms}.
The alignment of the characteristic extraction spacetime foliation with the
outgoing null rays and the choice of a convenient combination of independent
  Einstein field equation components ensures that the only Bondi-like
spin-weighted scalar that need be evolved between the
hypersurfaces is $J$.
The remaining metric quantities $\beta, U, Q, W$, and $H$ determined on each
hypersurface instead satisfy purely spatial partial differential equations,
which are hereafter and in the literature referred to as the set of
\emph{hypersurface equations}.

The Einstein field equations for the Bondi-like metric
(\ref{eq:BondiLikeMetric}) result in the following hypersurface equations, which
take a computationally convenient hierarchical form:
\begin{widetext}
\begin{subequations} \label{eq:BondiHierarchy}
  \begin{align}
\bl    \beta_{, r} &= S_{\bl \beta}(\bl J), \label{eq:BondiHierarchyBeta}\\
    (\bl r^2 \bl Q)_{,\bl r} &= S_{\bl Q}(\bl J,\bl \beta),\\
    U_{,\bl r} &= S_{\bl U}(\bl J,\bl \beta,\bl Q),\\
    (\bl r^2 \bl W)_{,\bl r} &= S_{\bl W}( \bl J,\bl \beta,\bl Q,\bl U),\\
    (\bl r \bl H)_{,\bl r} +  L_{\bl H}(\bl J,\bl \beta,\bl Q,\bl U,\bl W)\bl H
    + L_{\bl{\bar H}}(\bl J,\bl \beta,\bl Q,\bl U,\bl W) \bl{\bar{H}}
                &= S_{\bl H}(\bl J,\bl \beta,\bl Q,\bl U,\bl W).
  \end{align}
\end{subequations}
\end{widetext}

The form of the system (\ref{eq:BondiHierarchy}) allows for the technique of
first determining $\bl J$ on a hypersurface $\Sigma_{\bl u}$ of constant
$\bl u$, then using the value of $\bl J$ and the worldtube boundary value
$\beta|_{\Gamma}$ to determine $\beta$ on the hypersurface $\Sigma_{\bl u}$ via
(\ref{eq:BondiHierarchyBeta}), following the cascaded equations in order to
obtain $\bl H$.
Finally, $\partial_{\bl u} \bl J = \bl H$, so once $\bl H$ is determined by the
hypersurface equations, it may be used to step $\bl J$ to the next hypersurface.
Because $\partial_{\bl u} \bl J$ can be thus considered a (very complicated)
function of only $\bl J$, plus boundary data that depends on $\bl u$, the
stepping in $\bl u$ can be done using any ODE integration method.

The full form for each equation in (\ref{eq:BondiHierarchy}) may be found in a
variety of coordinate systems in prior derivations \cite{Bishop:1997ik,
  Handmer:2014qha} and is rederived in the accompanying Mathematica notebook
\cite{companion_package}.
However, practical implementations in spectral coordinates require an additional
coordinate transformation from previously published results, and the expressions
can be streamlined for numerical use by using identities between the Bondi-like
spin-weighted scalars.
In Section \ref{sec:compactified_evolution}, we present a simplified set of
hypersurface equations tailored to direct numerical implementation.

\subsection{Observables at $\mathcal{I}^+$} \label{sec:FoundationsScri}

The final stage of the previously developed CCE algorithm is to determine the
gravitational-wave observables from the Bondi-like spin-weighted scalars derived
from the evolution scheme described in the previous subsection
\ref{sec:EquationHierarchy}.
The primary gravitational-wave observable we consider is the Bondi news.
In a true Bondi-Sachs coordinate system (\ref{eq:BondiSachsMetric}), it is
defined as the leading part of the first time derivative of the angular metric:
\begin{equation} \label{eq:BondiNews}
  N_{\bs A \bs  B} = \lim_{\bs r\rightarrow \infty} \left(\bs r\partial_{\bs u} h_{\bs A \bs B} \right).
\end{equation}
When working with the spin-weighted metric quantities, the same information may
be conveyed by expressing the news as a spin-weighted scalar,
\begin{equation}
  N = \frac{1}{2}\bs{\bar{q}}^{\bs A} \bs{\bar{q}}^{\bs B} N_{\bs A \bs B}
  = \lim_{\bs r \rightarrow \infty} \left(\bs r \partial_{\bs u}  \bs J \right).
\end{equation}
Importantly, while the inputs to computing the Bondi-Sachs news $N$ depend on
the coordinate system, it is a gauge-invariant (up to the residual BMS freedom)
quantity defined such that in every coordinate system it takes the value
computed by the formula (\ref{eq:BondiNews}) in the Bondi-Sachs coordinates.
See \cite{Bishop:1997ik,Babiuc:2008qy} for alternative formulas for defining the
same Bondi news.

Previous treatments of the characteristic evolution systems then face the
challenge of computing the spin-weighted Bondi news function $N(u, x^{A})$ in a
Bondi-like frame by analytically computing a form for the news that holds for a
generic Bondi-like gauge \cite{Bishop:1997ik}.
That form of the news is then computed using the asymptotic values of the
spin-weighted scalars determined by the hypersurface equations
(\ref{eq:BondiHierarchy}).
However, the Bondi-like form of the news does not compensate for the fact that
the hypersurface equations have been solved in coordinates that are not
asymptotically inertial.

To account for the coordinate change at $\mathcal{I}^+$, the CCE algorithm then
computes the set of inertial coordinates
$\{\bs u(\bl u, \bl x^{\bl A}),\bs x^{\bs A}(\bl u, \bl x^{\bl A})\}$ on
$\mathcal{I}^+$ by solving the set of geodesic equations \cite{Bishop:1997ik},
\begin{subequations}\label{eq:GeodesicEquations}
  \begin{align}
    \partial_{\bl{u}} \bs x^{\bs A} &= - U^{\bl{B}} \partial_{\bl{B}} \bs x^{\bs A},\\
    \partial_{\bl{u}} \bs u &= \left( - U^{\bl{B}} \partial_{\bl{B}} \bs u + 1\right)e^{2 \bl \beta},
  \end{align}
\end{subequations}
which may be integrated in $u$ along $\mathcal{I}^+$.
Finally, the news function derived from Bondi-like quantities may be evaluated
in terms of the inertial coordinates $N(u, x^A)$.
This final result is then equivalent to the news that would have been computed
from the simpler form (\ref{eq:BondiNews}) if the system had been in a true
Bondi-Sachs gauge from the start.
The news in the inertial coordinates is then the primary gravitational
waveform observable provided by the CCE algorithm.

In previous work, the coordinate transformation accomplished by Eq.
~(\ref{eq:GeodesicEquations}) was considered only at $\mathcal{I}^+$.
In Section \ref{sec:bondi_transforms} below, we make use of a modified and
streamlined version of the same angular coordinate transformation to derive a
complete set of new transformations for the Bondi-like scalars throughout the
bulk of the spacetime.
Additionally, we extend the derivation to provide a set of new coordinate
conditions sufficient to place the bulk spacetime metric in the Bondi-Sachs
coordinates, which are then uniquely gauge-fixed up to BMS transformations.
Taken together with the prior results \cite{Bishop:1998uk} summarized in
\ref{sec:boundary_transforms}, the technique given in Section
\ref{sec:bondi_transforms} describes a concrete method to transform any input
metric to Bondi-Sachs coordinates.
Further, we derive an alternative form for the news function in an arbitrary
Bondi-like gauge that is compatible with previous results, but somewhat simpler
to compute.

The results of Section \ref{sec:bondi_transforms} below also grant the freedom
to apply the coordinate transformations at any point during the CCE algorithm.
We show in Section \ref{sec:regularity_preservation} that there is significant
numerical advantage to performing a portion of those coordinate transformations
on the worldtube boundary data \emph{before} the characteristic evolution.
The use of the worldtube coordinate transformation provably avoids pure-gauge
logarithmic dependence that has previously plagued spectral implementations of
CCE \cite{Handmer:2014qha}.

\section{Transformations from Bondi-like to Bondi
  coordinates} \label{sec:bondi_transforms}

In this section, we present a sequence of explicit coordinate transformations
that take any Bondi-like metric (\ref{eq:BondiLikeMetric}) to a true Bondi-Sachs
frame, in which the metric asymptotically approaches the Minkowski metric at the
appropriate rates determined in \cite{Bondi:1962px} and reviewed in Eqs.
~\eqref{eq:BondiSachsFalloffs}.
We refer to a new intermediate gauge derived in these steps as the
``partially flat'' gauge.
This gauge is well suited to numerical implementations and provably avoids
pure-gauge logarithms (see Section \ref{sec:regularity_preservation}).
Enforcing the partially flat gauge underlies many of the improvements we
derive in this paper for practical CCE implementations.

It is also worth noting that the transformations presented here are of general
interest to any computation for which it is desirable to determine a highly
fixed gauge from an input metric in arbitrary coordinates, such as self-force
calculations \cite{Campanelli:1998jv, Barack:2018yvs} or detailed comparisons
between general relativity computational techniques.
Starting from ADM quantities on a spherical 2-surface, the standard Cauchy
characteristic extraction descriptions from \cite{Bishop:1998uk} will determine
a metric in an arbitrary Bondi-like gauge, which is not yet unique up to BMS
transformations because of the significant freedom in asymptotic behavior.
The transformations in this section complete the procedure, giving a concrete
method to determine the true Bondi-Sachs metric, unique up to BMS
transformations, for the spacetime bulk.
We present the final transformation both as a partial differential equation that
may be solved for the exact Bondi-Sachs coordinates and as an expansion near
$\mathcal{I}^+$.

The coordinate transformations presented here not only improve the gauge of
practical computations, but also supply the metric components in a Bondi-Sachs
frame.
These metric components are easily related to asymptotic waveform quantities,
such as the news (\ref{eq:BondiNews}) and the Weyl scalars discussed in Section
\ref{sec:weyl_scalars} below.
Our form for the news is compatible with earlier computations of the same
quantity by other methods \cite{Bishop:1997ik, Babiuc:2008qy}.

\subsection{Step-by-step coordinate transformations} \label{sec:bondi_transform_steps}

For the detailed presentation of the coordinate transformations, it is 
convenient to work with the up-index form of the Bondi-like metric, as it more
easily maps to the transformations of the spin-weighted scalars
$\{\beta, W, U, J\}$.
Further, we adopt the radial coordinate $l = 1/r$, which is 
convenient for asymptotic expansions.
In cases where the only coordinate alteration is the use of the inverse radial
coordinate, we choose not to introduce new index embellishments to avoid
over-complicating the notation.
The up-index metric in the Bondi-like form (\ref{eq:BondiLikeMetric}) is
\begin{equation} \label{eq:UpIndexBondiLikeMetric}
  g^{\mu \nu} = \left[
  \begin{matrix}
      0 &  l^2 e^{-2 \beta} & 0 \\
       l^2 e^{-2 \beta} & l^3 e^{-2 \beta}  (W + l) &  l^2 e^{-2\beta} U^A\\
      0 & l^2 e^{-2 \beta} U^B & l^2 h^{A B}
    \end{matrix} \right].
\end{equation}

The procedure for completely establishing a volume Bondi-Sachs metric is
accomplished by the following sequence of steps.
\begin{enumerate}
\item Establish asymptotically inertial angular coordinates
  $\ifc{x}^{\ifc{A}}(u, x^A)$, removing the asymptotically constant part of
  $U^{A}$.
\item Modify the radial coordinate to $\ifc{l}(l, u, x^A)$ to enforce the
  desired determinant in the angular block of the metric.
 Following this step, the coordinate system is the partially flat
  coordinates $\{\ifc u = u, \ifc{l}, \ifc{x}^{\ifc{A}}\}$ we have referred to
  previously.
\item Alter the time coordinate to the asymptotically inertial $\bs{u}$,
  removing the asymptotically constant part of $\beta^{(0)}$.
 This will, however, disrupt the Bondi form of the metric, giving an
 $\mathcal{O}(\ifc{l}^2)$ contribution to $g^{\bs u \ifc A}$.
\item Impose the Bondi metric restriction $g^{\bs u \bs A} = 0$ by introducing a
  set of angular coordinates modified only at subleading order in $\ifc l$,
  $\bs{x}^{\bs{A}} = \ifc{x}^{\ifc{A}} + l \bs{x}^{(1)}(u, l, x^A)$.
\item Restore the Bondi-Sachs coordinate conditions by constructing an areal
  radius $\bs l$ associated with $\bs x^{\bs A}$, again modifying only the
  subleading part in $\ifc{l}$.
  Following this step, the coordinates $\{\bs u, \bs l, \bs x^{\bs A}\}$ are in the
  Bondi-Sachs frame, fixed up to BMS transformations.
\end{enumerate}
Following steps 2 and 5, we will present the metric in terms of the original
Bondi-like spin-weighted scalar components.
For simplicity of presentation, we do not do so for the metric following steps
1, 3, and 4, as the intermediate metric at those stages of the computation is
not even in Bondi-like form, so has limited utility.

We note that our set of coordinate transformations from steps 1 -- 5 is in
agreement with the set of inertial coordinates on $\mathcal{I}^+$ established by
\cite{Bishop:1997ik}, but possesses additional radial dependence necessary to
enforce an asymptotically flat Bondi metric that is also valid in the bulk of
the spacetime.
The conditions we impose are equivalent to the coordinate conditions previously
explicated for the Bondi-Sachs metric \cite{Barnich:2010eb}, but our derivation
gives rise to an important new intermediate gauge and produces explicit recipes
for the Bondi-Sachs quantities from arbitrary Bondi-like inputs.

Following the program laid out above, step 1 selects a set of angular
coordinates $\ifc{x}^{\ifc{A}}$ such that
\begin{equation} \label{eq:XIfcEom}
  \partial_u \ifc{x}^{\ifc{A}} = - \partial_A \ifc{x}^{\ifc{A}} U^{(0) A},
\end{equation}
where $U^{(0) A}$ is the asymptotic part of $U^A$ defined by the expansion
\begin{equation}\label{eq:AsymptoticExpansionOfU}
  U^A = U^{(0) A}(u, x^A) + l U^{(R) A}(u, l, x^A).
\end{equation}
In this expansion the superscript $(R)$ is used to indicate the
``remainder'' after the subtraction of the asymptotically constant $U^{(0)}$.
Note that (\ref{eq:XIfcEom}) can be numerically evaluated using standard
time-integration methods, and permits arbitrary initial angular coordinates
$\ifc{x}^{\ifc{A}}(u=0)$.
We choose a natural initial condition for this time-integration, which is to
take $x^{\ifc{A}}(u=0, x^A) = \delta^{\ifc{A}}{}_A x^A$.
Note that $l$ is no longer the areal radius for the new set of angular
coordinates.
Therefore, the intermediate metric between step 1 and step 2 is not Bondi-like,
so could not be used in a characteristic evolution without re-deriving the
equations of motion.

In step 2 we restore Bondi-like form by performing a transformation to an areal
inverse radius $\ifc{l}$ for the new set of angular coordinates
$\ifc{x}^{\ifc{A}}$.
The new areal coordinate is determined by the alteration of the angular
determinant:
\begin{equation} \label{eq:conformal_factor_determinant}
  \ifc{l} = l \sqrt{\det(\partial_A \ifc{x}^{\ifc{A}})}
\left(  \frac{\det(\ifc{q}_{\ifc{A} \ifc{B}})}{\det(q_{A B})}\right)^{1/4}
  \equiv \frac{l}{\ifc \omega(u, x^A)}.
\end{equation}
To more simply express the transformations of the spin-weighted scalars
following this coordinate transformation, we introduce the spin-weighted
Jacobian factors,
\begin{subequations}\label{eq:SpinWeightedJacobiansAB}
\begin{align}
  \ifc a = \ifc q^{\ifc A} \partial_{\ifc A} x^{A} q_{A},\\
  \ifc b = \ifc{\bar{q}}^{\ifc A} \partial_{\ifc A} x^{A} q_{A}.
\end{align}
\end{subequations}
Under these definitions, we have the convenient identity for the conformal factor $\ifc \omega$:
\begin{equation} \label{eq:ifc_omega}
  \ifc \omega = \frac{1}{2}\sqrt{\ifc b \ifc{\bar{b}} - \ifc a \ifc{\bar{a}}}.
\end{equation}

After steps 1 and 2, the new
spin-weighted scalar components of the metric are:
\begin{subequations} \label{eq:ifcSwScalars}
  \begin{align}
    \ifc \beta =& \beta - \frac{1}{2} \log \ifc \omega \label{eq:ifcBeta},\\
    \ifc J =& \frac{\ifc{\bar b}^2 J + \ifc a^2 \bar J + 2 \ifc a \ifc{\bar b} K}{4 \ifc \omega^2},\\
    \ifc U =& \frac{1}{2\ifc \omega^2} \left(\ifc{\bar b} (U - U^{(0)}) - \ifc a (\bar U - \bar U^{(0)})\right) \notag\\
                &- \frac{\ifc l e^{2 \ifc \beta}}{\ifc \omega} \left(\ifc \eth \ifc \omega \ifc K - \ifc{\bar{\eth}} \ifc \omega \ifc J\right),\\
    \ifc W =& W + (\ifc \omega - 1) \ifc l - \frac{2 \partial_{\ifc u} \ifc \omega}{\ifc \omega}- \frac{1}{\ifc \omega} \left(\ifc U \ifc {\bar \eth} \ifc \omega+ \ifc{\bar U} \ifc \eth \ifc \omega\right) \notag\\&+ \frac{e^{2 \ifc \beta} \ifc l}{2 \ifc \omega^2}\left(\ifc{\bar \eth} \ifc \omega^2 \ifc J+ \ifc \eth \ifc \omega^2 \ifc{\bar{J}}- 2 \ifc \eth \ifc \omega \ifc{\bar \eth} \ifc \omega \ifc K\right)  \label{eq:ifcW}.
  \end{align}
\end{subequations}
The hypersurface equations for $W$ and $\hat{W}$ or expansion of derivatives in
the determinant (\ref{eq:conformal_factor_determinant}) may be used to infer
$\partial_{\ifc u} \ifc \omega$:
\begin{subequations}
\begin{align}
  \partial_{\ifc u} \ifc \omega =& \frac{\ifc \omega}{4}(\ifc \eth\,\bar{ \mathcal{U}}^{(0)}+ \ifc{\bar{\eth}} \, \mathcal{U}^{(0)} )\notag\\&+ \frac{1}{2} \left(\mathcal U^{(0)} \ifc{\bar{\eth}} \ifc \omega+ \bar{\mathcal{U}}^{(0)} \ifc \eth \omega\right)\\
  \mathcal{U}^{(0)} \equiv& \frac{1}{2 \ifc \omega^2} \left(\ifc{\bar b} U^{(0)} - \ifc a \bar U^{(0)} \right).
\end{align}
\end{subequations}
Our choice to use angular coordinates $\ifc x^{\ifc A}$ that initially coincide
with the Cauchy angular coordinates $x^A$ ensures that
$\ifc \omega(u=0,x^A) = 1$.
At later times, the conformal factor is calculated from the angular derivatives
of the new angular coordinates via Eq.~(\ref{eq:ifc_omega}).
The differentiation of the formulas (\ref{eq:ifcSwScalars}) to obtain $\ifc Q$
and $\ifc H$ in the new coordinates is nontrivial and critical for computational
steps of the regularity-preserving scheme described in Section
\ref{sec:regularity_preservation}, so we give those pieces explicitly in
Appendix \ref{sec:IF_extras}.
The reader may refer to the metric forms (\ref{eq:UpIndexBondiLikeMetric}) and
(\ref{eq:BondiLikeMetric}) for re-assembling the coordinate components of the
metric.

At this point, the intermediate metric after step 2 already has some attractive
properties.
The metric is again Bondi-like, and the new $\ifc{U}^{\ifc{A}}$ vanishes at
$\mathcal{I}^+$, which offers a number of simplifications to other metric
components via the Einstein equations.
Once we have determined the corresponding $\ifc{U}^{\ifc{A}}$,
$\ifc{h}^{\ifc{A} \ifc{B}}$, $\ifc{\beta}$, and $\ifc{W}$, we may also
make use of the existing formalism for evolving the Bondi-like system in the
bulk.
This coordinate system forms the foundation of a regularity-preserving
evolution, in which the volume Cauchy Characteristic evolution is conducted in
the coordinates $\{\ifc u, \ifc{l}, \ifc{x}^{\ifc{A}}\}$, and the worldtube
quantities are transformed to these coordinates before integration.
In this coordinate system, if $\ifc{h}^{(0) \ifc{A} \ifc{B}}(u=0) = q^{A B}$,
then $\ifc{h}^{(0) \ifc{A} \ifc{B}} = q^{A B}$ for the entire evolution, where
$\ifc{h}^{(0) \ifc{A} \ifc{B}}$ is the asymptotic part of
$\ifc{h}^{\ifc{A} \ifc{B}}$ in the same way that $U^{(0)A}$ was defined by Eq.
~(\ref{eq:AsymptoticExpansionOfU}).
In terms of the spin-weighted quantity $\ifc{J}$, this is equivalent to saying
that if $\ifc{J}$ vanishes at $\mathcal{I}^+$ on the initial data hypersurface,
it will vanish at $\mathcal{I}^+$ for the entire evolution.
We call this set of intermediate coordinates the \emph{partially flat}
coordinates.

However, the discussion so far has not yet arrived at an asymptotically flat
metric, so we return to our transformation sequence with step 3, which
introduces the time coordinate $\bs{u}$ such that
\begin{equation}
  \bs{u} = \bs u^{(0)}(\ifc u, \ifc{x}^{\ifc A}) +
  \ifc{l} \bs u^{(R)}(\ifc u, \ifc{l}, \ifc{x}^{\ifc{A}})
  = \int^{\ifc u} e^{2 \ifc \beta} 
  + \ifc{l} \bs u^{(R)}(\ifc u, \ifc{l}, \ifc{x}^{\ifc{A}}).
\end{equation}
We fix $\bs u^{(R)}$ by insisting that the resulting metric has vanishing
$g^{\bs{u} \bs{u}}$,
\begin{widetext}
\begin{align} \label{eq:uRequation}
  0 =   g^{\bs u \bs u} &= 2 \left(e^{2 \ifc \beta}  + \ifc l \partial_{\ifc u} \bs u^{(R)} \right) \, \partial_{\ifc l} \left(\ifc{l} \bs u^{(R)}\right) \ifc l^2 e^{-2 \ifc \beta}  + \ifc l^3  \partial_{\ifc l}(\ifc l \bs u^{(R)}) ^2 (\ifc W + \ifc l) e^{-2 \ifc \beta}\notag \\
  & +  2 \ifc l^2 \left(\partial_{\ifc{A}} \bs u^{(0)}
     + \ifc l \partial_{\ifc A} \bs u^{(R)}\right)
    \partial_{\ifc l} \left(\ifc l \bs u^{(R)}\right) \ifc U^{\ifc A} e^{-2 \ifc \beta}
    + \ifc{l}^2 (\partial_{\ifc A} \bs u^{(0)} + \ifc l \partial_{\ifc A} \bs u^{(R)}) (\partial_{\ifc B} \bs u^{(0)} + \ifc l \partial_{\ifc B} \bs u^{(R)}) \ifc h^{\ifc A \ifc B},
\end{align}
\end{widetext}
which specifies $u^{(R)}$ as the solution to an elliptic equation.
We note two possible strategies for solving (\ref{eq:uRequation}).
First, the expansion of $u^{(R)}$ in ascending powers of $\ifc{l}$ gives rise to
algebraic solutions for each successive expansion coefficient.
For illustration, we explicitly compute the first two orders of
$\bs u^{(R)} =\bs u^{(1)} + \ifc l\bs u^{(2)} + \mathcal{O}(\ifc l^2) $ in
Appendix \ref{app:perturbative_transformations}.
Fortunately, for the discussion of asymptotic quantities in this paper, this
power series solution is all that is necessary.
Second, it may be possible to solve Eq.~(\ref{eq:uRequation})
for $\partial_{\ifc{l}} (\ifc{l} u^{(R)})$ and
numerically step $u^{(R)}$ along the hypersurface given some single sphere of
values for $u^{(R)}$, though we make no claims here of the stability of such a
numerical scheme.

After step 3 we are again left with a metric that is not Bondi-like, because of
nonvanishing $g^{u \ifc{A}}$.
In the transformation step 4, we make a subleading in $\ifc l$ change to the
angular coordinates, affecting only the values of $\bs{x}^{\bs{A}}$ away from
$\mathcal{I}^+$.
The new angular coordinates take the form
\begin{equation}
  \bs{x}^{\bs{A}} = \delta^{\bs{A}}{}_{\ifc{A}} x^{\ifc{A}}
  + \ifc{l} \bs{x}^{(R) \bs{A}} (\ifc{u}, \ifc{l}, \ifc{x}^{\ifc{A}}).
\end{equation}
We constrain $\bs x^{\bs A}$ in much the same way we constrained $\bs u$: We
impose $g^{\bs u \bs A} = 0$ to preserve the Bondi form.
This condition gives the equation
\begin{widetext}
\begin{align}\label{eq:bsxA}
  0 = g^{\bs u \bs A} =& \partial_{\ifc u} \bs u\, \partial_{\ifc l}(\ifc l \bs x^{(R) \bs A}) \ifc l^2 e^{-2 \ifc \beta} + \partial_{\ifc l} \bs u \,\partial_{\ifc u} \bs x^{(R) \bs A} \ifc l^3 e^{-2\ifc \beta} + \partial_{\ifc l} \bs u\, \partial_{\ifc l} (\ifc l \bs x^{(R) \bs A}) \ifc l^3 (\ifc W + \ifc l) e^{-2 \ifc \beta}\notag\\
  &+ \ifc l^2 (\delta^{\bs A}{}_{\ifc A} + \ifc l \partial_{\ifc A} \bs x^{(R) \bs A})\, \partial_{\ifc l} \bs u\,\ifc U^{\ifc A} e^{-2 \ifc \beta} + \ifc l^2 \partial_{\ifc l}(\ifc l \bs x^{\bs A}) \partial_{\ifc A} \bs u \,\ifc U^{\ifc A} e^{-2 \ifc \beta} + \ifc l^2 \partial_{\ifc B} \bs u (\delta_{\ifc A}{}^{\bs A} + \ifc l \partial_{\ifc A} \bs x^{(R) \bs A}) h^{\ifc A \ifc B}.
\end{align}
\end{widetext}
For small $\ifc{l}$, (\ref{eq:bsxA}) may be perturbatively expanded to
obtain an algebraic equation for each order of $\ifc{x}^{(R) \ifc{A}}$, and
otherwise must be numerically integrated along $\hat{l}$.
The first two orders of the asymptotic expansion are provided in Appendix
\ref{app:perturbative_transformations}.

The final coordinate transformation step 5 is closely analogous to step 2,
except now we are restoring the areal radius associated with the angular
coordinates $\bs x^{\bs A}$.
Therefore, the Bondi-Sachs radial coordinate is
\begin{align}\label{eq:Step5RadialCoordinate}
    \bs{l} &= \ifc l \sqrt{\det(\partial_{\ifc A} \bs{x}^{\bs{A}})}
  \left(\frac{\det(\bs{q}_{\bs{A} \bs{B}})}{\det(q_{\ifc A \ifc B})}\right)^{1/4}\notag\\
  &\equiv \frac{\ifc l}{\bs \omega(\ifc u, \ifc l, \ifc x^{\ifc A})} \equiv \ifc{l} - \ifc{l}^2 \bs \omega^{(1)} + \mathcal{O}(\ifc{l}^3).
\end{align}

In a manner analogous to Eqs.~(\ref{eq:SpinWeightedJacobiansAB}) and~(\ref{eq:ifc_omega}), $\bs \omega$ can be written in terms of the derivatives
  $\partial_{\bs A} \ifc{x}^{\ifc{A}}$.  These derivatives are determined by solving Eq.~(\ref{eq:bsxA}) perturbatively, and doing so yields a perturbative solution for $\bs \omega$.  The expression for $\bs \omega^{(1)}$ is provided
  in Appendix
\ref{app:perturbative_transformations}.

The final formulas for the Bondi-Sachs spin-weighted scalars in the spacetime
volume after all of these coordinate transformations are comparatively
uninformative.
Because the subleading transformations in steps 3 -- 5 depend on all of the
partially flat coordinates, all 4 partially flat spin-weighted scalars
contribute to each of the Bondi-Sachs scalars according to the standard metric
transformation rules.
However, the final transformations do provide compact results under perturbative
expansion near $\mathcal{I}^+$:
\begin{subequations}
  \begin{align}
    \bs \beta^{(0)} &= \bs \beta^{(1)} =  \bs J^{(0)} = 0 \\
    \bs U^{(0)} &= \bs U^{(1)} = \bs W^{(0)} = \bs W^{(1)} = 0,\\
    \bs J^{(1)} &= \ifc J^{(1)} + \ifc \eth^2 \bs u^{(0)}.
  \end{align}
\end{subequations}
In some of these equations, we have further simplified using results from
the perturbative expansion of the Einstein field equations.

\subsection{Inference of the news in arbitrary coordinate systems} \label{sec:news}

The Bondi news is a gauge invariant measure of the gravitational wave signal:
the ${}\sim 1/r$ part of the angular Bondi-Sachs metric is uniquely specified by
the Bondi-Sachs metric and falloff properties.
Because this gives a unique gauge specification (up to BMS freedom), the gauge-invariant Bondi-Sachs news is defined as taking the value \emph{in every
  gauge} that is given by transforming to the Bondi-Sachs gauge and evaluating
\begin{equation}\label{eq:NewsDefinitionBondi}
  N = \lim_{\bs r \rightarrow \infty} \frac{1}{2 \bs r} \bs{\bar{q}}^{\bs{A}} \bs{\bar{q}}^{\bs{B}} \partial_{\bs{u}}\bs h_{\bs A \bs B}.
\end{equation}
Therefore, the explicit coordinate transformation given in Section
\ref{sec:bondi_transform_steps} gives a novel route to a gauge-invariant
specification of the Bondi-Sachs news in terms of the components of a Bondi-like
metric (\ref{eq:BondiLikeMetric}).

For our recommended numerical technique of evolving the characteristic system of
equations in the partially flat coordinates, we determine the gauge invariant
Bondi-Sachs news by expanding (\ref{eq:NewsDefinitionBondi}) in terms of
partially flat quantities via expressions from stages 3 -- 5 of the
transformations in \ref{sec:bondi_transform_steps}. The Bondi-Sachs news in
terms of quantities in the partially flat gauge is
\begin{equation}\label{eq:NewsDefinitionIncompletelyFlat}
  N = e^{-2 \ifc \beta^{(0)}} \left(\ifc{\bar H}^{(1)} + \ifc{\bar \eth} \ifc{\bar \eth} e^{2 \ifc \beta^{(0)}}\right).
\end{equation}

In addition, we derive an equation for the Bondi-Sachs news in terms of
spin-weighted scalars in an arbitrary Bondi-like gauge, which may be of use to
computations that use strategies other than our partially flat coordinate
system.
This definition of the news is equivalent to previous expansions of the same
quantity \cite{Bishop:1997ik}, but our set of explicit coordinate
transformations yields simpler expressions than previously derived formulas.
Our simplified expression is obtained by using the relationships between the
derivatives $\ifc \eth$ and the Bondi-like $\eth$, and the transformation of
$\ifc H$ (\ref{eq:ifcH}) and $\ifc \beta$ (\ref{eq:ifcBeta}), and takes the form
\begin{widetext}
\begin{align}\label{eq:NewsDefinitionBondiLike}
  N =& \frac{\ifc \omega^2 e^{-2 \beta^{(0)}}}{4} \bigg\{
2  \left[
\ifc{\bar b} \eth \bar U^{(0)}  J^{(1)} + \ifc a \eth U^{(0)} \bar J^{(1)} + \left(\ifc{\bar b} \eth U^{(0)} + \ifc a \eth \bar U^{(0)}\right)\text{Re}\left(J^{(0)} \bar{J}^{(1)} \right)
       \right]\notag\\
     &+\left[\ifc{\bar b}^2 H^{(1)} + \ifc a^2 \bar H^{(1)} + \ifc{\bar b} \ifc a\left(2 \text{Re}\left(J^{(0)} \bar{H}^{(1)}  + J^{(1)} \bar H^{(0)}\right) - \text{Re}\left(J^{(0)} \bar{H}^{(0)}\right) \text{Re}\left(J^{(0)} \bar J^{(1)}\right) \right)\right] \notag\\
     &+ \frac{1}{2} \left[ \ifc{\bar b}^2 \left(U^{(0)} \bar \eth + \bar U^{(0)} \eth\right) J^{(1)} + \ifc a^2 \left(U^{(0)} \bar \eth + \bar U^{(0)} \eth\right) \bar{J}^{(1)} + \ifc a \ifc{\bar b} \left(U^{(0)} \bar \eth + \bar U^{(0)} \eth\right) \text{Re}\left(J^{(0)} \bar J^{(1)}\right) \right]\notag\\
&+  3\ifc \omega^2 \partial_u \ifc \omega \left[\ifc{\bar b} J^{(1)} + \ifc a^2 \bar{J}^{(1)} + \ifc a \ifc{\bar b} \text{Re}\left(J^{(0)} \bar J^{(1)}\right)\right] \
       \bigg\} +  \frac{\ifc \omega e^{-2 \beta^{(0)}}}{4} \left(\ifc b^2 \eth^2  + \ifc{\bar a}^2 \bar \eth^2 + 2 \ifc b \ifc{\bar a} \eth \bar \eth \right)\left(\frac{e^{2 \beta^{(0)}}}{\ifc \omega}\right).
\end{align}
\end{widetext}

\section{Compactified characteristic evolution equations} \label{sec:compactified_evolution}

In this section, we present the full set of nonlinear characteristic equations
using a compactified radial coordinate.
We start with any coordinate system $\{u,r,\theta,\phi\}$ where the metric is in
Bondi-like form, Eq.~(\ref{eq:BondiLikeMetric}), and obeys the Bondi-like
restrictions, Eqs~(\ref{eq:BondiLikeFalloff}).
Note that this includes more restricted coordinates such as the partially
flat coordinates of Section~\ref{sec:bondi_transform_steps} and true Bondi-Sachs
coordinates, so the results of this section are applicable to those more
restricted coordinates as well.

The characteristic equations presented here are equivalent to a
coordinate-transformed version of the equations derived by \cite{Bishop:1997ik},
but take a newly updated form.
The new compactified radial coordinate gives the left-hand sides of the
differential equations a standard and convenient form, and because it is is
defined on $[-1, 1]$ it facilitates standard spectral tools such as Legendre or
Chebyshev polynomial representations.
In addition to introducing a compactified coordinate, we have also manipulated
the equations in the following ways to better suit numerical implementation:
\begin{enumerate}
\item All spin-weighted derivatives result in a spin-weight between 2 and $-2$,
  so that spin-weighted transform libraries with a hard limit at spin weight 2
  can be used.
\item All derivatives of $K$ are expanded and, to the extent possible,
  simplified with other terms involving $J$.
\item Where relevant, terms that nearly cancel have been reduced to expressions
  of similar numerical magnitude.
 For example, consider the term $1-K$.
If evaluated directly for small $J$, there will be loss of numerical precision.
 In the extreme case where $J$ is on the order of the square root of machine
 epsilon, then $K$ will be unity to machine precision and $1-K$ will be entirely
 error.
Our solution is to evaluate $1-K$ as $(1-K^2)/(1+K) = J\bar{J}/(1+K)$, which
 retains full precision even when $J$ is small.
\item For applicable equations, many collections of terms that appear along with
  their complex conjugates are identified so that several numerical steps can be
  avoided by caching a quantity and negating its imaginary part.
\end{enumerate}

Given any Bondi-like (including even more restricted) coordinates
$\{u,r,\theta,\phi\}$, we define compactified coordinates
$\{\na{u}, \na{y}, \na{\theta}, \na{\phi}\}$.
The numerical coordinates are identical to the Bondi-like coordinates aside from
the radial component,
\begin{equation}
\na  u = \bl{u},\qquad \na y = 1 - \frac{2 R}{\bl{r}},
  \qquad \na \theta = \bl{\theta}, \qquad \na \phi = \bl{\phi},
\end{equation}
where $R$ is the Bondi-like radius of the worldtube $\Gamma$,
\begin{equation}
  R(u, \theta, \phi) = \bl{r}|_{\Gamma}.
\end{equation}
By construction, $\Gamma$ is a surface of constant $y = -1$,
but not a surface of constant Bondi-like radius $r$.

Inverting the Jacobian gives the necessary factors to convert between the
Bondi-like derivatives and the derivatives with respect to the numerical
coordinates:
\begin{subequations} \label{eq:numerical-jacobian}
\begin{align}
  \partial_{\bl r} =& \frac{(1 - \na y)^2}{2 R} \partial_{ \na y},\\
  \partial_{\bl u} =& \partial_{\na u} - (1 - \na y) \frac{\partial_{\na u} R}{R} \partial_{\na y}, \label{eq:numerical-jacobian-u} \\
  \bl  \eth =& \na \eth - (1 - \na y) \frac{\na \eth R}{R} \partial_{\na y}.
\end{align}
\end{subequations}

In the following discussion, there is often little to be gained from explicity
converting the angular derivatives $\eth$ and $\bar{\eth}$ to the numerical
coordinates $\na \eth$ and $\na{\bar \eth}$ in the analytic equations.
In virtually every case, the numerical angular derivatives do not simplify the
expression.
Further, new terms introduced by the angular Jacobians are better behaved at
$\mathcal{I}^+$, so the statements about the pole structure of the equations
carry through without modification.
In cases where $\bl \eth$ and $\bl \eth$ act on functions of $\na{x}^{\na \mu}$,
the needed Jacobian factors from (\ref{eq:numerical-jacobian}) are implied.
For a practical implementation, the appropriate Jacobian factors should
be included in the subroutines that calculate the angular derivatives $\eth$ and
$\bar{\eth}$.
However, it is valuable to convert explicit factors of $r$ and derivatives with
respect to $r$ to numerical coordinates, as such factors can alter the pole
structure in the numerical coordinates and offer subtle simplifications.

The results of this section are the new explicit forms of the equations
themselves, so are presented in full in below subsections
\ref{sec:compactified-beta}-\ref{sec:compactified-H} with minimal further
exposition on their structure. The reader is invited to download the Mathematica
companion notebook to this paper \cite{companion_package}, which provides tools
for confirming the compactified characteristic equations and their digital
forms for further exploration. The equations are given in terms of
Bondi-like spin-weighted scalars $\na J, \na \beta, \na Q, \na U, \na W,$ and
$ \na H$ in the compactified coordinates. Most of these scalars depend on the
coordinate derivatives only through the equations of motion, so the quantities
act as scalars for the transformation $\na F = \bl F$, for all
$F \in \{J, \beta, Q, U, W\}$, and
\begin{equation} \label{eq:numerical-h-conversion}
  \na H = H + \partial_u R \partial_r J.
\end{equation}

\subsection{Common forms of hypersurface equations}

 Each of the
hypersurface equations (\ref{eq:BondiHierarchy})
 can be placed into one of three categories, depending
 on the pole structure of the equation
 (arising from terms like $(1-\na y)^n$ that vanish at $\mathcal{I}^+$)
 and the form of the terms that involve the variable
  that is determined by that equation.
 We discuss these categories in terms
of our compactified radial coordinate $\na y$, but most statements here would be
similarly applicable to any choice of radial coordinate.
The first and simplest type is those hypersurface equations
 that govern the
quantities $\beta$ and $U$. These have the form
\begin{equation} \label{eq:form1}
  \partial_{\na y} F_1(\na x^{\na \mu}) = S_1(\na x^{\na \mu}),
\end{equation}
where $F_1$ is the quantity determined by the equation, and $S_1$ is a nonlinear
expression that is independent of $F_1$ but may involve other already-known
hypersurface quantities.
For this category, the integral determining $F_1$ can be evaluated in a
straightforward way by standard ODE methods.

The second category contains the hypersurface equations that
govern the computation of $Q$ and $W$.
These possess the form
\begin{align} \label{eq:form2}
  (1- \na y)& \partial_{\na y} F_2(\na x^{\na \mu})
  + 2 F_2(\na x^{\na \mu})\notag\\
  &= S_2^{P}(\na x^{\na \mu}) + (1 - \na y) S_2^{R}(\na x^{\na{\mu}}).
\end{align}
Here $F_2$ is the quantity being solved for, and $S_2^{P}$ and $S_2^{R}$ are
known nonlinear expressions independent of $F_2$.
This form of equation requires significantly more care than Eq.~(\ref{eq:form1})
because of the $(1 - \na y)$ terms.
Depending on the method of solution, erroneous logarithm dependence may arise
from manipulation of the differential equation \cite{Barkett:2019uae}.
Further, the known functions $S_2^{P}$ and $S_2^{R}$ on the right-hand side of
Eq.~\eqref{eq:form2} produce a quadratic contribution when they are expanded as a
power series in $(1 - \na y)$ near $\na y = 1$.
Unless there is an explicit cancellation of this quadratic in $(1 - \na y)$
contribution, this equation produces pure-gauge logarithm dependence in the
solution to $F_2$.
In Section \ref{sec:bondi_transforms} we present a set of coordinate
transformations that we use in Section \ref{sec:regularity_preservation} to
construct a complete method for avoiding the logarithm dependence in these
contributions to the Bondi-like characteristic equations.

The third and final form of differential equation found among the characteristic
hypersurface equations governs only the quantity $H$. It takes the form
\begin{widetext}
\begin{equation} \label{eq:form3}
  (1 - \na y) \partial_{\na y} F_3(\na{x}^{\na \mu}) + \big[1 + (1 - \na y)L_3^G(\na x^{\na \mu}) L_3^J(\na x^{\na \mu})\big] F_3(\na x^{\na \mu}) + (1-\na y)\bar L_3^G(\na x^{\na \mu}) L_3^J(\na x^{\na \mu}) \bar{F}_3(\na x^{\na \mu}) = S_3^P(\na x^{\na \mu}) + (1-\na y)S_3^R(\na x^{\na \mu}).
\end{equation}
\end{widetext}
Here $F_3$ is the quantity being solved for, and $S_3^P$, $S_3^R$, $L_3^G$,
and $L_3^J$ are known nonlinear expressions independent of $F_3$.
The treatment of the poles for this equation is similar to that used for
 Eq.~(\ref{eq:form2}),  except
that it is now the linear in
$(1-\na y)$ term that governs the logarithmic dependence.
The form (\ref{eq:form3}) has the further complication of the non-derivative
term depending both on $F_3$ and its complex conjugate $\bar{F}_3$, which
prevents a simple single-pass integration in spectral representation that is
possible for (\ref{eq:form1}) and (\ref{eq:form2}).
Instead, we decompose the equation and the individual
factors into real and imaginary parts.
The matrix form of Eq.~(\ref{eq:form3}) is
\cite{Handmer:2014qha}
\begin{widetext}
\begin{align}
\left(\left[\begin{array}{cc}(1-y)\partial_y + 1 & 0\\ 0 & (1-y)\partial_y + 1\end{array} \right] +  (1 - y)  \left[\begin{array}{cc} \text{Re}(L_3^J)\text{Re}(L_3^G) &  \text{Re}(L_3^J) \text{Im}(L_3^G)\\
          \text{Im}(L_3^J) \text{Re}(L_3^G) &  \text{Im}(L_3^J) \text{Im}(L_3^G)\end{array}
      \right]\right)\left[\begin{array}{c}\text{Re}(F_3)\\\text{Im}(F_3)\end{array}\right] \notag\\= \left[\begin{array}{c}\text{Re}(S_3^P) + (1 - y) \text{Re}(S_3^R)\\ \text{Im}(S_3^P) + (1 - y) \text{Im}(S_3^R)\end{array}\right].
\end{align}
\end{widetext}  
\subsection{Hypersurface equation: $\beta$} \label{sec:compactified-beta}

The first equation to be evaluated on each hypersurface is the one that
determines $\beta$ given $J$ on the same hypersurface:
\begin{align} \label{eq:HypersurfaceBeta}
\partial_r \beta = - \frac{1}{8} r \left(\partial_r J \partial_r \bar{J} -  \frac{\left(\partial_r (J \bar{J})\right)^2}{4 K^2}\right).
\end{align}
Converting to numerical coordinates  yields
\begin{align} \label{eq:Betanumeric}
\partial_{\na y} (\na \beta) = -\frac{1}{8} (1 - {\na y}) \left(\partial_{\na y} \na J \partial_{\na y} \na{\bar{J}} -  \frac{ \left(\partial_{\na y} (\na J \na{\bar{J}})\right)^2}{4 \na K^2}\right).
\end{align}
This equation takes the form (\ref{eq:form1}) and can be integrated by
traditional methods.


\subsection{Hypersurface equation: $Q$}

The second hypersurface equation evaluated on each hypersurface determines the
value of $Q$ given $J$ and $\beta$.
Our simplified form of the $Q$ hypersurface equation in Bondi-like coordinates
is
\begin{align} \label{eq:HypersurfaceQ}
\partial_r (Q r^2) =& - r^2 \left(\Lambda_Q + \frac{\bar{\Lambda}_Q J}{K} + \frac{\partial_r \bar{\eth} J}{K}\right)\notag\\& + 2 r^4 \partial_r \left(\frac{\eth \beta}{r^2}\right),
\end{align}
where
\begin{align}
  \Lambda_Q =& - \tfrac{1}{2} \eth (\bar{J} \partial_r J) + \tfrac{1}{2} J \partial_r \eth \bar{J} \notag\\
  &-  \tfrac{1}{2} \eth \bar{J} \partial_r J + \frac{\eth (J \bar{J}) \partial_r (J \bar{J})}{4 K^2}.
\end{align}
These equations become, when converted to compactified coordinates,
\begin{widetext}
\begin{align} \label{eq:Qnumeric}
  2 \na Q + (1 -  {\na y}) \partial_{\na y} \na Q  = -4 \eth  \na \beta -  (1 -  {\na y})\left(2 \Lambda_{\na Q} + \frac{2 \bar{\Lambda}_{\na Q} \na J}{\na K} - 2  \eth \partial_{\na y} \na \beta + \frac{\bar{\eth} \partial_{\na y} \na J}{\na K} - \frac{2 \eth R \partial_{\na y} \na \beta}{R} + \frac{\bar{\eth} R \partial_{\na y} \na J}{R \na K}\right),
\end{align}
where we have introduced
\begin{align}
\Lambda_{\na Q} = - \tfrac{1}{4} \eth (\na{\bar{J}} \partial_{\na y} \na J) + \tfrac{1}{4} \na J \eth \partial_{\na y} \na{\bar{J}} -  \tfrac{1}{4} \eth \na{\bar{J}} \partial_{\na y} \na J + \frac{\eth (\na J \na{\bar{J}})   \partial_{\na y} (\na J \na{\bar{J}} )}{8 K^2} + \frac{\eth^\prime (R)\left(\na J \partial_{\na y} \na{\bar{J}} -\na{\bar{J}} \partial_{\na y} \na J\right)}{4 R}.
\end{align}
\end{widetext}
The pole structure is easily inferred from the form of equation
(\ref{eq:Qnumeric}).
We identify the first and second terms of the right-hand side with $S^P_{\na Q}$
and $S^R_{\na Q}$ from (\ref{eq:form2}).

\subsection{Hypersurface equation: $U$}

The next equation in the sequence determines the value of $U$ (analogous to an
angular shift contribution in the ADM formalism) on a given hypersurface given
the values of $J$, $\beta$, and $Q$.
In Bondi-like coordinates
\begin{equation} \label{eq:HypersurfaceU}
\partial_r U = \frac{e^{2 \beta}}{r^2} \left(K Q - J \bar{Q}\right).
\end{equation}
The $U$ equation is of unusual simplicity for the characteristic formulation, so
the conversion to our adjusted compactified form is straightforward:
\begin{equation} \label{eq:Unumeric}
  \partial_{\na y} \na U = \frac{e^{2 \na \beta}}{2 R} \left(\na K \na Q - \na J \na{\bar{Q}}\right).
\end{equation}
Similar to the equation for $\na \beta$, the hypersurface equation for $\na U$
is of the form (\ref{eq:form1}) and can be integrated by traditional means.

\subsection{Hypersurface equation: $W$} \label{sec:compactified-W}

The fourth equation in the hypersurface integration sequence determines the
value of the ``mass aspect'' contribution $W$ on a $u=\text{constant}$ hypersurface
given values of $J, \beta, Q$, and $U$:
\begin{align} \label{eq:HypersurfaceW}
  \partial_r (r^2 W) =& 1 +  \tfrac{1}{2} e^{2 \beta} (\Lambda_W + \Lambda_W) + (\eth \bar{U} + \bar{\eth} U) r\notag\\
  & + \tfrac{1}{4} (\partial_r \eth \bar{U} + \partial_r \bar{\eth} U) r^2,
\end{align}
where
\begin{widetext}
\begin{align}
\Lambda_W =& - \eth \beta \eth \bar{J} + \tfrac{1}{2} \bar{\eth} \bar{\eth} J + 2 \bar{\eth} \beta \bar{\eth} J + (\bar{\eth} \beta)^2 J + \bar{\eth} \bar{\eth} \beta J + \frac{\eth (J \bar{J}) \bar{\eth} (J \bar{J})}{8 K^3} + \frac{2 + J \bar J}{2 K} -  \frac{\eth \bar{\eth} (J \bar{J})}{8 K} \nonumber \\ &-  \frac{\eth (J \bar{J}) \bar{\eth} \beta}{2 K} 
-  \frac{\eth \bar{J} \bar{\eth} J}{4 K} -  \frac{\bar J \eth \bar{\eth} J}{4 K}   -  K \eth \bar{\eth} \beta  - K  \eth \beta \bar{\eth} \beta + \tfrac{1}{4} (- K Q \bar{Q} + J \bar{Q}^2).
\end{align}

The conversion to numerical coordinates for this equation proceeds simply, as no
term in $\Lambda_W$ requires any alteration (replace $\beta$, $J$, $K$, and $Q$ with $\na \beta$, $\na J$, $\na K$, and $\na Q$).
We re-arrange to make explicit the pole structure of the equation,
\begin{align} \label{eq:Wnumeric}
  2 \na W& + (1 - \na y) \partial_{\na y} \na W \notag\\
         & = \left(\eth \na{\bar{U}} + \bar{\eth} \na U\right) + (1 -\na  y)\left(\tfrac{1}{4} \eth \partial_{\na y} \na{\bar{U}} + \tfrac{1}{4} \bar{\eth} \partial_{\na y} \na U  + \frac{1}{4} \partial_{\na y} \na U \frac{\bar{\eth} R}{R} + \frac{1}{4} \partial_{\na y} \na{\bar{U}} \frac{\eth R}{R}-  \frac{1}{2 R} + \frac{e^{2 \na \beta} (\Lambda_{\na W} + \bar{\Lambda}_{\na W})}{4 R}\right).
\end{align}
Again drawing the comparison to (\ref{eq:form2}), we identify the first and
second terms on the right-hand side of (\ref{eq:Wnumeric}) as $S^P_{\na W}$ and
$S^R_{\na W}$.
\end{widetext}

\subsection{Hypersurface equation: H} \label{sec:compactified-H}

The final hypersurface equation determines the value of
$\partial_{u} J \equiv H$, which is the sole time-derivative quantity used in
the characteristic evolution computation.
Once determined by radial integration, the numerical time derivative
$\partial_{\na u} J$ is used in a standard numerical ODE integrator (e.g.,
Runge-Kutta or adaptive Dormand-Prince) to determine the value of $J$ on
subsequent hypersurfaces. This value
is then used as the input for the hypersurface
computations of the metric components on subsequent hypersurfaces.

The simplified hypersurface equation for $H$ in Bondi-like coordinates is
\begin{widetext}
\begin{align} \label{eq:HypersurfaceH}
&\partial_r  (H r) + J r (\mathcal{L}_H H + \bar{\mathcal{L}}_H H )\notag\\ &= - \frac{1}{2} \eth ((J + r \partial_r J ) \bar{U}) + (\mathcal{B}_H + \bar{\mathcal{B}}_H) J - J\left( \eth \bar U + \frac{1}{2} \bar \eth U\right) -  K \eth U -  \frac{1}{2} \partial_r (r \bar{\eth} J ) U + \frac{1}{2} \partial_r \partial_r J (r^2 W + r)  \notag\\
&\hspace{.5cm}+ \frac{e^{2 \beta}}{r} \left[\frac{1}{2} \bar{\eth} J K \mathcal{C}_H  + \frac{\bar{\mathcal{C}}_H  J^2 \eth \bar J}{2 K}  -  \left(\mathcal{A}_H + \bar{\mathcal{A}_H} +  \frac{\eth \bar{\mathcal{C}_H}}{2 K}\right) J + \eth \mathcal C_H - \frac{\eth (J \bar{\mathcal{C}}_H)}{2 K}  + \mathcal{C}_H^2\right] \nonumber \\ 
& \hspace{.5cm}+ \partial_r (J) \left[1 + \frac{1}{2} r^2 \partial_r W + \frac{3}{2} r W - \frac{1}{2} \left(- \bar{\eth} \bar{U} J + \frac{\eth U \bar{J}}{K^2}\right) K r - \frac{1}{4} \mathcal D_H  K^2 r\right]\notag\\
&\hspace{.5cm} + \partial_r (\bar{J}) J^2 r \left[ \frac{1}{4} \mathcal D_H  + \frac{1}{2 K}\eth U  \bar{J} \right],
\end{align}
where
\begin{subequations}
\begin{align}
  \mathcal{A}_H =& \frac{1}{4} \eth (\eth (\bar{J})) -  \frac{\eth \bar{\eth} (J \bar{J}) -  2 \bar J \eth \bar{\eth} J }{16 K^3} -  \frac{\bar J \eth \bar{\eth} J - 3}{4 K} + \frac{1}{2} \eth (\bar{J}) \left(\mathcal C_H + \frac{\bar{\eth} (J \bar{J}) J}{4 K^3} -  \frac{\bar{\eth} J (2 J \bar J + 1)}{4 K^3} \right),\\
  \mathcal{B}_H =&\frac{1}{2 r} +  W + \frac{1}{2} r \partial_r W + 2 \partial_r \beta (r W + 1)\notag\\
                 &-  \frac{\eth U \bar{J} \partial_r (J \bar{J}) r}{4 K} + \frac{r U}{4} \left(\bar{\eth} J \partial_r \bar{J}  + \bar \eth (\bar J \partial_r J)  -  \bar J \partial_r \bar \eth J - \frac{\bar{\eth} (J \bar{J}) \partial_r (J \bar{J})}{2 K^2}\right)  \label{eq:script-b} ,\\
  \mathcal{C}_H =&  \eth \beta -  \frac{1}{2} Q,\\
  \mathcal{D}_H =& \eth \bar U - \bar \eth U,\\
\mathcal{L}_H =& \frac{1}{2} \left(- \partial_r \bar{J} + \frac{\bar{J} \partial_r (J \bar{J})}{2 K^2}\right).
\end{align}
\end{subequations}
\end{widetext}
The equation for $H$ is considerably more intricate than the other hypersurface
equations.
When writing it in numerical coordinates, note that we must also rewrite the
evolution equation $\partial_{u}J = H$ as $\partial_{\na u}\na J = \na H$, where
$\na H$ is related to $H$ by Eq.
~(\ref{eq:numerical-h-conversion}).
The hypersurface equation for $\na H$ takes the form (\ref{eq:form3})
\begin{widetext}
\begin{equation} \label{eq:Hnumeric}
  (1 - \na y) \partial_{\na y} \na H +\na H + (1 - \na y)\na J (\mathcal{L}_{\na H} \na H
  + \bar{\mathcal{L}}_{\na H} \na{\bar{H}}) = S_{\na H}^P  + (1 - \na y) S_{\na H}^R,
\end{equation}
with source terms
\begin{subequations} \label{eq:numerical-evolution-source-terms}
  \begin{align}
    \mathcal{L}_{\na H} =& \frac{1}{2} \left(- \partial_{\na y} \na{\bar J} + \frac{\na{\bar J} \partial_{\na y}(\na J \na{\bar J} )}{2 \na K^2}\right),\\
    S_{\na H}^P =& - \frac{1}{2} \left(\eth (\na J \na{\bar U}) +  \bar \eth(\na J \na U) + 2 J \eth \bar U\right) + 2\na W \na J  - \na K  \eth \na U,\\
    S_{\na H}^R =& \na J (\mathcal{B}_{\na H} + \bar{\mathcal{B}}_{\na H}) + \frac{1}{2} (1-\na y) \left(\na W + \frac{(1-\na y)}{2 R} + 2 \frac{\partial_{\na u} R}{R}\right) \partial_{\na y}^2 \na J - \frac{1}{2} \left(\na U \bar \eth \partial_{\na y} \na J + \na {\bar U} \eth \partial_{\na y} \na J\right) +\na J^2 \partial_{\na y} \na{ \bar J} \left[\frac{\mathcal{D}_{\na H}}{4} + \frac{\eth \na U \na{\bar J}}{2 \na K}\right]\notag\\
                    & + \frac{\partial_{\na y} \na J}{2} \left[- \frac{1}{2}(\eth \bar{\na U} + \bar \eth \na U)- \na U \frac{\bar \eth R}{R} - \na{\bar U} \frac{\eth R}{R} + \na W + (1- \na y) \partial_{\na y} \na W - \na K \left(\frac{\eth \na U \na{\bar J}}{\na K} - \bar \eth \na{\bar U} \na J\right) - \frac{1}{2}(\na K^2 + 1) \mathcal{D}_{\na H} - \eth \na{\bar U} \partial_{\na y} \na J\right]\notag\\
    & + \frac{e^{2 \na \beta}}{2 R} \left[ \frac{J}{2 K^3} + \frac{1}{2} \bar{\eth}\na J \na K \mathcal{C}_{\na H}  + \frac{\bar{\mathcal{C}}_{\na H} \na J^2 \eth \na{\bar J}}{2 \na K}  -  \left(\mathcal{A}_{\na H} + \bar{\mathcal{A}}_{\na H}\right)\na J + \eth \mathcal C_{\na H} - \frac{ \bar{\mathcal{C}}_{\na H} \eth \na J}{2\na K}  + \mathcal{C}_{\na H}^2\right],
  \end{align}
\end{subequations}
where $\mathcal{A}_{\na H}$, $\mathcal{C}_{\na H}$, and $\mathcal{D}_{\na H}$
have the same form as $\mathcal{A}_H$, $\mathcal{C}_H$, and $\mathcal{D}_{H}$
with the replacements
$\{J,\beta,Q,U,W\}\rightarrow \{\na J, \na \beta, \na Q, \na U, \na W\}$.
The quantity $\mathcal{B}_{\na H}$ is defined as
\begin{align}
  \mathcal{B}_{\na H} = \frac{1}{2} \bigg[& \frac{1}{2 R} + \partial_{\na y} \na W + \left(\na W + \frac{1 - \na y}{2 R} + 2 \frac{\partial_{\na u} R}{R}\right) \partial_{\na y} \na \beta - \frac{\eth \na U \na{\bar J} \partial_{\na y} (\na J \na{\bar J})}{2 \na K} \notag\\
                                          &+ \frac{\na U}{2} \left(\bar \eth \na J \partial_{\na y} \na{\bar J} + \bar \eth(\na{\bar J} \partial_{\na y} \na J) - \na{\bar J} \bar \eth \partial_{\na y} \na J - \frac{\bar \eth (\na J \na{\bar J}) \partial_{\na y} (\na J \na{\bar J})}{2 \na K^2}\right)\bigg]
\end{align}
\end{widetext}

\section{Regularity-preserving CCE}
\label{sec:regularity_preservation}

One of the most notable drawbacks of previous implementations
\cite{Handmer:2014qha} of spectral CCE was the presence of pure-gauge
logarithmic dependence, where spin-weighted scalars like $Q$, $W$, and $H$
developed behavior like $r^{-n} \ln(r)$ at large $r$.
The logarithmic dependence is particularly troubling for spectral treatments of
the system, which rely on the representability of all solutions as a rapidly
converging polynomial series in the compactified radial coordinate.
This problem was partially mitigated in the updated implementation
\cite{Barkett:2019uae}, which ensures that no logarithmic behavior is introduced
by inadequate numerical treatment of the equations if such behavior is not
present in the true solutions of those equations.
However, there is no guarantee that the true solutions of the characteristic
equations lack all logarithmic dependence in an arbitrary Bondi-like gauge.

In this section, we present the necessary conditions for the CCE system to
remain regular (all quantities are polynomial in $r^{-1}$)
at $\mathcal{I}^+$ as the system is
evolved. 
We demonstrate that the intermediate gauge, referred to here as the partially
flat gauge, is sufficient to guarantee asymptotically well-behaved spin-weighted
scalars throughout the evolution.
We then synthesize the collection of our suggested improvements in a complete
description of an improved Cauchy-characteristic evolution algorithm in Section
\ref{sec:computation_roadmap}.

\subsection{An overview of the regularity conditions in abstract notation}
\label{sec:regularity_preservation_overview}

The hypersurface equations for $Q$, $W$, and $H$ each have a nontrivial pole
structure.
In this section, we explore that pole structure, and the consequences of the
partially flat gauge developed in Section \ref{sec:bondi_transforms} for the
regularity of $\ifc Q$, $\ifc W$, and $\ifc H$.
To avoid complications with order counting for powers of $r$ when derivatives
are involved, we adopt the frequently used notation $l \equiv r^{-1}$ to
describe asymptotic dependence, where $l= 0$ at $\mathcal{I}^+$.
Note also that the reasoning presented in this section applies similarly to the
numerical coordinates, using $\na l = (1- \na y)$.
For simplicity of notation, we present the regularity conditions in the
Bondi-like coordinates, and note that if the spin-weighted scalars are regular
in Bondi-like $l$ at $\mathcal{I}^+$, they are also regular in the associated
$(1 - \na y)$ numerical coordinate.

The $Q$ and $W$ equations take the form (\ref{eq:form2}).
Expressed in terms of $l$, we have the form
\begin{equation} \label{eq:QWabstract}
\partial_l \left(\frac{F_2}{l^2}\right) = \frac{S_2^P}{l^3} + \frac{S^R_2}{l^2}
\end{equation}
For generic right-hand side factors $S_2^P$ and $S^R_2$, which depend on $l$,
there is the danger that the right-hand side asymptotically scales as $l^{-1}$.
If this happens, either initially or during an evolution, the solution for $Q$
or $W$ will behave asymptotically as $l^2 \ln(l)$.
The conditions to avoid such logarithmically dependent terms in these equations
are the \emph{regularity conditions} on the hypersurface sources for the
characteristic equations.
For the above equation (\ref{eq:QWabstract}), the regularity condition is
\begin{align}\label{eq:QWregularity}
  \frac{1}{2} \partial_{l}^2 S_2^P |_{l = 0} &= - \partial_l S^R_2|_{l = 0} 
\end{align}

Similar conditions apply to the hypersurface equation that
determines the
evolution quantity $\partial_u J \equiv H$ (c.f. (\ref{eq:form3})),
which we write in terms of $l$ as
\begin{equation} \label{eq:Habstract}
  \partial_l \left(\frac{F_3}{l}\right) + \frac{1}{l}( L_{F_3} F_3 + L_{\bar F_3} \bar F_3) = \frac{S_3^P}{l^2} + \frac{S_3^R}{l}.
\end{equation}
The regularity condition for Eq.~(\ref{eq:Habstract}) is
\begin{equation} \label{eq:Hregularity}
  \partial_l S_3^P |_{l = 0} = - S_3^R |_{l = 0}.
\end{equation}
Note that because $H=\partial_u J$, $H$ must also obey any requirement imposed
on $J$ by the regularity conditions (\ref{eq:QWregularity}) and
(\ref{eq:Hregularity}); otherwise, $J$ will fail to obey the regularity
requirements as it evolves forward in time.

\subsection{Explicit form of the regularity conditions}
\label{sec:regularity_preservation_conditions}

Here, we present the conclusions regarding the regularity of the set of
hypersurface and evolution equations obtained by perturbatively expanding each
equation near $\mathcal{I}^+$.
In each step, we fix the first few powers in $l$ of the spin-weighted scalar in
question needed for subsequent steps, identify components informed by boundary
data, and determine the constraints implied by the regularity conditions
(\ref{eq:QWregularity}) and (\ref{eq:Hregularity}).
In particular, we show that the partially flat gauge established in Section
\ref{sec:bondi_transforms} is sufficient to ensure regularity of the
hypersurface equations when supplied with appropriate initial data for $\ifc J$.
It is important to note that an partially flat gauge is not necessary, and
more complicated gauge conditions can be imposed to ensure regularity in more
generic Bondi-like gauges.
In the below expansions, we formally impose the partially flat requirements
$\ifc J^{(0)} =\ifc U^{(0)} = 0$.
All other results follow from these requirements used in conjunction with the
hypersurface equations from Section \ref{sec:compactified_evolution} and the
regularity conditions (\ref{eq:QWregularity}) and (\ref{eq:Hregularity}).
In this section, we use expansion notation consistent with
Section~\ref{sec:bondi_transforms}; for instance,
$\beta = \beta^{(0)} + l \beta^{(1)} + l^2 \beta^{(2)} + \mathcal{O}(l^3)$.

The $\ifc \beta$ hypersurface equation (\ref{eq:HypersurfaceBeta}) gives rise to
no regularity conditions; if $\ifc J$ is regular at $\mathcal{I}^+$, then so is
$\ifc \beta$.
Expanding the generic $\ifc \beta$ hypersurface equation order-by-order assuming
$\ifc J$ regularity, we conclude that for an partially flat gauge,
\begin{subequations} \label{eq:BetaNearScri}
\begin{align}
\ifc  \beta^{(1)} &= 0,\\
\ifc  \beta^{(2)} &= \frac{\ifc{\bar J}^{(1)} \ifc J^{(1)}}{16},
\end{align}
\end{subequations}
and $\ifc \beta^{(0)}$ is fixed by the boundary conditions.

The $\ifc Q$ hypersurface equation (\ref{eq:HypersurfaceQ}) gives rise to the
requirements
\begin{subequations}
  \begin{align}
    \ifc Q^{(0)} &= - 2 \ifc \eth \ifc \beta^{(0)},\\
    \ifc Q^{(1)} &= \ifc{\bar \eth} \ifc J^{(1)}.\label{eq:SubleadingQRegularity}
  \end{align}
\end{subequations}
So, the $\mathcal{O}(l)$ part of the $Q$ hypersurface equation
(\ref{eq:SubleadingQRegularity}) gives rise to the regularity condition
\begin{equation}
\ifc  J^{(2)} = 0,
\end{equation}
and the $\ifc Q^{(2)}$ part of the expansion is determined by the boundary data on the hypersurface.

Regularity of $\ifc U$ follows from regularity of the previously discussed $Q$,
$J$, and $\beta$.
However, for subsequent equations, we require the results from the
perturbative expansion of the $\ifc U$ hypersurface equation
(\ref{eq:HypersurfaceU}):
\begin{subequations}
  \begin{align}
    \ifc U^{(1)} &= 2 e^{2\ifc \beta^{(0)}} \ifc \eth \ifc \beta^{(0)},\\
    \ifc U^{(2)} &= -\frac{e^{2 \ifc \beta^{(0)}} (\ifc{\bar \eth} \ifc J^{(1)} + 2 \ifc{\bar \eth} \ifc \beta^{(0)} \ifc J^{(1)})}{2}.
  \end{align}
\end{subequations}

From the expansion of the $\ifc W$ hypersurface equation
(\ref{eq:HypersurfaceW}), we have the order-by-order constraints:
\begin{subequations} \label{eq:WNearScri}
  \begin{align}
    \ifc W^{(0)} &= 0,\\
    \ifc W^{(1)} &= e^{2 \ifc \beta^{(0)}} - 1 + 2 e^{2 \ifc \beta^{(0)}} \ifc \eth \ifc{\bar \eth} \ifc \beta^{(0)} + 4 e^{2 \ifc \beta^{(0)}} \ifc \eth \ifc \beta^{(0)} \ifc{\bar \eth} \ifc \beta^{(0)}.
  \end{align}
\end{subequations}
The regularity of $\ifc W$ is directly satisfied from the previous identities,
so it imposes no further conditions.

Finally, we consider the equation (\ref{eq:HypersurfaceH}) for $\ifc H$.
The regularity condition again demands that the ${}\sim l$ part of the right-hand
side vanishes, which is already satisfied given the conditions of the
partially flat gauge and the results
(\ref{eq:BetaNearScri}--\ref{eq:WNearScri}).
The expansion of $\ifc H$ also fixes
\begin{subequations} \label{eq:EvolutionNearScri}
  \begin{align}
    \partial_{\ifc u}\ifc J^{(0)} = 0,\\
    \partial_{\ifc u} \ifc J^{(2)} = 0,
  \end{align}
\end{subequations}
and $\partial_{\ifc u} \ifc J^{(1)}$ is fixed by boundary conditions.
The constraints given by (\ref{eq:EvolutionNearScri}) are important, as they
indicate that the previous requirements constructed to ensure regularity are
stable in the evolution system.

We conclude from the full examination of the characteristic system of equations
that to ensure regularity for the entire evolution, it is sufficient to impose
the two conditions:
\begin{itemize}
\item Partially flat gauge
\item $\ifc J^{(2)} = 0$ on the initial hypersurface.
\end{itemize}
Without these requirements, worldtube data provided in an arbitrary Bondi-like
gauge will tend to have small violations of the regularity conditions and
produce slowly growing logarithmic terms that undermine the precision of
spectral techniques.

\subsection{Computational gauge strategy} \label{sec:computation_roadmap}

In the above Subsections
\ref{sec:regularity_preservation_overview}--\ref{sec:regularity_preservation_conditions},
we've demonstrated that under a particular set of asymptotic flatness
requirements, the characteristic evolution system remains regular and possesses
desirable computational properties.
However, in standard CCE implementations, the Bondi-like coordinates are subject
to the arbitrary gauge provided by the worldtube data, and it is not immediately
obvious how we might practically implement the transformations necessary to keep
the evolution system in the partially flat gauge.

In this section, we present a sketch of the computational strategy for imposing
the partially flat gauge given the typical starting point of unfixed data
on the initial hypersurface $u=0$ and boundary data provided in a Bondi-like
gauge at the worldtube.
The procedure we outline below demonstrates that it is possible to obtain all of
the information necessary to impose regularity and partially flat gauge
without disrupting the hierarchical structure of the characteristic system of
equations (\ref{eq:BondiHierarchy}).

\vspace{2mm}

{\bf Initialization:}
\begin{enumerate}
\item Initialize the partially flat angular coordinates $\ifc{x}^{\ifc{A}}$ on the initial hypersurface $u=0$
  as the Bondi-like angular coordinates $\ifc{x}^{\ifc A}(u=0) = x^A(u=0)$.
\item Initialize $\ifc{J}$ by using the provided boundary value of $J$ (for the
  initial time the angular transformation is trivial), with a $r^{-1}$ falloff.
 Further terms may be added to adhere more tightly to the boundary data, but no
  term should be added that has either $r^{0}$ or $r^{-2}$ dependence near
  $\mathcal{I}^+$.
\end{enumerate}

At the start of the evolution, we know $\partial_A \ifc x^{\ifc A}$ at the
current $u$, but we do \emph{not} know $\partial_u \ifc x^{\ifc A}$, as the time
derivative requires knowledge of $U^{(0)}$ (cf.~Eq.~\ref{eq:XIfcEom}).

\vspace{2mm}
 
{\bf Evolution:}
\begin{enumerate}
\item Compute $\ifc{\beta}|_{\Gamma}$ on the worldtube using the conversion
  equation (\ref{eq:ifcBeta}) and interpolating to the grid associated with the new
  angular coordinates $\ifc x^{\ifc A}$, then
  evaluate the $\ifc{\beta}$ hypersurface equation (\ref{eq:Betanumeric}), taking as
  input $\ifc \beta|_{\Gamma}$ and $\ifc J$.
\item Compute $\ifc Q|_{\Gamma}$ on the worldtube using the transformation
  equation (\ref{eq:ifcQ}) and interpolating to $\ifc x^{\ifc A}$, then evaluate
  the $\ifc{Q}$ hypersurface equation (\ref{eq:Qnumeric}) using the
  boundary value $\ifc Q|_\Gamma$ and the previously determined $\ifc \beta$ and
  $\ifc J$.
\item The $U$ equation requires care.
 Note that at this point, we do not know $U^{(0)}$ at the current timestep, so
  we must find a method of determining both $U^{(0)}$ and $\ifc{U}$ given what
  we have so far calculated.
 To accomplish this, we make use of
  \begin{equation} \label{eq:UHat}
    \mathcal{U} = \frac{1}{2 \hat{\omega}^2} \left(  \ifc{\bar b} U - \ifc a \bar{U} \right) - \frac{\ifc le^{2 \ifc \beta}}{\ifc \omega} \left(\ifc \eth \ifc \omega \ifc K - \ifc{\bar \eth} \ifc \omega \ifc J\right).
  \end{equation}
 Note that $\mathcal{U} - \hat{U} = - (1/2 \hat{\omega}^2) (\ifc{\bar b} U^{(0)} - \ifc a \bar{U}^{(0)})$, so
  $\partial_r \mathcal U = \partial_r \ifc U$, so we may use as the integrand for
  $\mathcal{U}$ the right-hand side of (\ref{eq:Unumeric}), taking as inputs the previously calculated Bondi quantities
  $\ifc{J}, \ifc{\beta},$ and $\ifc{Q}$.
 Then, to integrate $\mathcal{U}$, we evaluate (\ref{eq:UHat}) on the boundary and
 integrate the right-hand side evaluated from partially flat quantities $\ifc J$,
 $\ifc \beta$, and $\ifc Q$.
\item Determine the $\mathcal I^+$ value  $U^{(0)}$ from the $\mathcal{I}^+$ value of
  $\mathcal{U}$ determined in the previous step.
 This involves only the mild inconvenience of the complex matrix inversion, giving
 \begin{equation}
   U^{(0)} = \frac{1}{2 \ifc \omega^2} \left(\bar b\, \mathcal{U}^{(0)}
     - a\, \mathcal{\bar{U}}^{(0)}\right).
 \end{equation}
 We have now established the rest of the coordinate transformation Jacobian
 \begin{equation}
   \partial_u \ifc{x}^{\ifc{A}} = - U^{(0) B} \partial_B  \ifc{x}^{\ifc{A}},
 \end{equation}
 So the remaining steps of the characteristic evolution proceed comparatively
 directly.
\item Evaluate $\ifc{U} = \mathcal{U} - \mathcal{U}^{(0)}$, which is the
  required spin-weighted scalar in the partially flat gauge to be used in the
  remainder of the calculation
\item The boundary condition for $\ifc{W}|_\Gamma$ is given by the
  transformation (\ref{eq:ifcW}).
 After determining the boundary value, evaluate the $\ifc{W}$ hypersurface
  equation (\ref{eq:Wnumeric}) using the previously calculated hypersurface
  values in the partially flat coordinates $\ifc J$, $\ifc \beta$, $\ifc Q$,
  and $\ifc U$.
\item Determine the partially flat boundary data $\ifc H|_\Gamma$ using
  (\ref{eq:ifcH}), then evaluate the $\ifc{H}$ hypersurface equation (\ref{eq:Hnumeric}) using
  the previously calculated $\ifc J$, $\ifc \beta$, $\ifc Q$, $\ifc U$, and
  $\ifc W$.
\item Evolve $\ifc{J}$ using
  $\ifc{H} \equiv \partial_{\ifc{u}} \ifc{J}$
\item Evolve $\ifc{x}^{\ifc{A}}(u)$ using
  $\partial_u \ifc{x}^{\ifc{A}} = - U^{(0) B} \partial_B
  \ifc{x}^{\ifc{A}}$
\item Iterate to subsequent timesteps starting at item 1 of the evolution procedure.
\end{enumerate}

\section{Newman-Penrose Weyl scalars from Bondi or Bondi-like
  metric} \label{sec:weyl_scalars}

Having determined the metric throughout the compactified null region with the
CCE algorithm described in Section \ref{sec:regularity_preservation} above, we
are now in a position to compute gauge-invariant (up to BMS) quantities that
describe the dynamics of the spacetime at $\mathcal{I}^+$.
In this section, we give an explicit dictionary between the complementary
Bondi-Sachs and Newman-Penrose formalisms.
Then, we take advantage of the relation between the asymptotic Weyl scalars in
the Newman-Penrose formalism and the spin-weighted scalars from the Bondi-like
metric to obtain simple expressions for the asymptotically leading contributions
to the Weyl scalars in our partially flat gauge.
The result is a simple prescription for adding the full set of Weyl scalars to
the outputs of the CCE algorithm.

\subsection{Newman-Penrose tetrad for Bondi-like coordinates}

To calculate the spin coefficients and Weyl scalars, we must select a reasonable
choice of orthonormal null tetrad.
We adopt a tetrad motivated by \cite{Babiuc:2008qy}, but modified with phase
($m^\mu \rightarrow e^{i \delta} m^\mu$) and overall factor
($n^\mu\rightarrow A n^\mu$ and $l^\mu \rightarrow l^\mu/A$) to asymptotically
match tetrads frequently used in numerical relativity \cite{Boyle:2019kee}:
\begin{subequations} \label{eq:NP_tetrads}
  \begin{align}
    m^\mu &= -\frac{1}{\sqrt{2} r}\left(\sqrt{\frac{K + 1}{2}}q^\mu
    - \sqrt{\frac{1}{2(1 + K)}} J \bar{q}^\mu\right),\\
    n^\mu &= \sqrt{2} e^{-2 \beta} \left(\delta^\mu{}_u - \frac{V}{2 r} \delta^\mu{}_r
            + \frac{1}{2} \bar{U}  q^\mu + \frac{1}{2} U \bar{q}^\mu\right),\\
    l^\mu &= \frac{1}{\sqrt{2}} \delta^\mu{}_r.
  \end{align}
\end{subequations}
We emphasize that while these tetrads asymptotically match those often used
  in numerical relativity, this does \emph{not} mean that all of the
  Newman-Penrose spin coefficients and Weyl scalars will similarly match.
 The subleading in powers of $1/r$ corrections to the tetrads can cause subtle
  alterations to the scalar functions, so care must be taken to ensure
  consistent conventions when comparing calculations.
 More details about the variety of Newman-Penrose conventions in numerical
  calculations can be found in \cite{DanteInPrep}. Here the spin-weighted scalars
$\beta$, $U$, $V$, $J$, and $K$ are assumed to be in generic Bondi-like
coordinates, but the results of this section apply to more restrictive
coordinates like Bondi-Sachs or partially flat
coordinates; in more restrictive coordinates many of the results simplify
further.
The tetrads (\ref{eq:NP_tetrads}) satisfy the standard normalization and
orthogonality conditions
\begin{subequations} \label{eq:NP_orthonormal}
\begin{align}
  -l^\mu n_\mu &= m^\mu \bar{m}_\mu = 1\\
  l^\mu l_\mu = l^\mu m_\mu &= n^\mu n_\mu = n^\mu m_\mu = m^\mu m_\mu = 0.
\end{align}
\end{subequations}
From (\ref{eq:NP_orthonormal}), we may write the metric as a combination of these vectors
\begin{equation}
  g_{\mu \nu} = - 2 l_{(\mu} n_{\nu)} + 2 m_{(\mu} \bar{m}_{\nu)}.
\end{equation}
In addition, it is convenient to define a set of scalar covariant derivatives in
terms of the Newman-Penrose null tetrads:
\begin{align}\label{eq:NPCovariantDerivatives}
  D = l^\mu \nabla_\mu, && \Delta = n^\mu \nabla_\mu, && \delta = m^\mu \nabla_\mu.
\end{align}
 
\subsection{Concise closed-form Newman-Penrose spin coefficients}

The Newman-Penrose covariant analogues of the independent Christoffel symbol
components are known as \emph{spin coefficients}, and are defined in terms of
the orthonormal null tetrad:
{\allowdisplaybreaks
 \begin{subequations}
  \begin{align}
    \kappa &=- m^\mu l^\nu \nabla_\nu l_\nu,\\
     \rho &= -m^\mu \bar{m}^\nu \nabla_\nu l_\mu,\\
    \sigma &=- m^\mu m^\nu \nabla_\nu l_\mu,\\
     \tau &= -m^\mu n^\nu \nabla_\nu l_\mu,\\
    \nu &= \bar{m}^\mu n^\nu \nabla_\nu n_\mu,\\
     \mu &= \bar{m}^\mu m^\nu \nabla_\nu n_\mu,\\
    \lambda &= \bar{m}^\mu \bar{m}^\nu \nabla_\nu n_\mu,\\
     \pi &= \bar{m}^\mu l^\nu \nabla_\nu n_\mu,\\
    \epsilon &= \frac{1}{2} (\bar{m}^\mu l^\nu \nabla_\nu m_\mu
     - n^\mu l^\nu \nabla_\nu l_\mu), \\
    \beta_{\text{NP}} &=  \frac{1}{2} (\bar{m}^\mu m^\nu \nabla_\nu m_\mu
      - n^\mu m^\nu \nabla_\nu l_\mu), \\
    \gamma &= \frac{1}{2} (\bar{m}^\mu n^\nu \nabla_\nu m_\mu
     - n^\mu n^\nu \nabla_\nu l_\mu),\\
    \alpha &= \frac{1}{2} (\bar{m}^\mu \bar{m}^\nu \nabla_\nu m_\mu
      - n^\mu \bar{m}^\nu \nabla_\nu l_\mu).
  \end{align}
\end{subequations}}
To resolve the notation collision between the Bondi-like $\beta$ that appears in
the metric (\ref{eq:BondiSachsMetric}) and the Newman-Penrose spin coefficient
that is traditionally notated with the same symbol, we write the spin
coefficient with a subscript: $\beta_{\text{NP}}$.

It is also convenient to define the unit spherical connection coefficient
$\Theta = q^A \bar q{}_B \nabla_A q^B{}$.
Under these definitions, the resulting spin-weighted Bondi-like expressions for
the Newman-Penrose spin coefficients are
\begin{widetext}
{\allowdisplaybreaks
\begin{subequations} \label{eq:spin_coefficients}
  \begin{align}
    \kappa &= 0,\\
    \rho &=- \frac{1}{\sqrt{2} r},\\
    \sigma &= -\frac{(1 + K) \partial_r J}{4 \sqrt{2} K}
              +  \frac{J^2 \partial_r \bar{J}}{4 \sqrt{2} K(1 + K)}, \\
    \tau &= -\frac{(2 \bar{\eth} \beta - \bar Q) J }{4 r \sqrt{1 + K}}
            +  \frac{(2\eth (\beta) - Q) \sqrt{1 + K}}{4 r},\\
    \nu &= \frac{\eth W  \bar{J}}
           {2 e^{2 \beta} \sqrt{1 + K} }
           - \frac{\bar{\eth} W \sqrt{1 + K}}
           {2 e^{2 \beta} },\\
    \mu &=\frac{e^{-2 \beta}(\eth \bar U + \bar \eth U)}{2 \sqrt{2}} - \frac{e^{-2\beta} (r^2 W + r)}{\sqrt{2} r^2},\\
    \lambda &= \frac{e^{-2 \beta}}{\sqrt{2}}\bigg[
               \frac{1}{2} \bar J \left(\bar \eth U - \eth \bar U\right) - \frac{\bar J^2 \eth U}{2(1 + K)} + \frac{\bar \eth \bar U (1 + K)}{2}  + \frac{r^2 W + r}{ r} \left(\frac{\bar J^2 \partial_r J}{1 + K}-\partial_r \bar J (1 + K)\right) \notag\\
         &\hspace{1.5cm}+ \frac{U}{4 K} \left(\bar \eth \bar J (1+K) -\frac{\bar J^2 \bar \eth J}{1 + K}\right)
           + \frac{\bar U}{4 K} \left( \eth \bar J (1 + K) - \frac{\bar J^2 \eth J}{1 + K}\right) - \frac{\bar J ^2 \partial_u J}{2 K (1 + K)} + \frac{(1 + K)\partial_u \bar J}{2K}\bigg],
    \\
    \pi &=  \frac{(2 \eth \beta +  Q)\bar J }{4 r \sqrt{1 + K}}
            -  \frac{(2\bar \eth (\beta) +\bar Q) \sqrt{1 + K}}{4 r},\\
    \epsilon &=  \frac{\partial_r (\beta)}{\sqrt 2}
                 + \frac{J \partial_r (\bar{J}) - \bar{J} \partial_r (J)}{8 \sqrt 2 (1 + K)}, \\
    \beta_{\text{NP}} &= \frac{1}{4 r \sqrt{1 + K}} \bigg[\bar \eth \beta J - \eth \beta (1 + K) -\frac{\eth (J \bar J)}{4 K} - \frac{J(\bar \eth (J \bar J) - \bar J \bar \eth J)}{4 K (1 + K)}  + \frac{\bar \eth J (1 + 3 K)}{4 K} \notag\\
               &\hspace{2.5cm}+ \frac{1}{2} J \bar Q - \frac{1}{2} Q (1 + K) - \frac{1}{2} J \bar{\Theta} - \frac{1}{2}(1 + K) \Theta \bigg],\\
    \alpha &=  \frac{1}{4 r \sqrt{1 + K}} \bigg[ \eth \beta \bar J - \bar \eth \beta (1 + K) +\frac{\bar \eth (J \bar J)}{4 K} + \frac{\bar J( \eth (J \bar J) -  J  \eth \bar J)}{4 K (1 + K)}  - \frac{\eth J (1 + 3 K)}{4 K} \notag\\
               &\hspace{2.5cm}+ \frac{1}{2} \bar J  Q - \frac{1}{2} \bar Q (1 + K) + \frac{1}{2} \bar J \Theta + \frac{1}{2}(1 + K) \bar \Theta \bigg],\\
    \gamma &= \frac{e^{-2 \beta}}{\sqrt{2}}\bigg[
               \frac{1}{4} \left(J \bar \eth \bar U - \bar J \eth U\right) + \frac{K}{4}\left(\bar \eth U - \eth \bar U\right)  + \frac{r^2 W + r}{8r(1 + K)} \left(\bar J \partial_r J - J \partial_r \bar J\right) + \frac{1}{2}(r \partial_r W +  W) \notag\\
         &\hspace{1.5cm}+ \frac{U\left(\bar \eth(J \bar J) - 2 \bar J \bar \eth J\right)}{8 (1 +K)} 
           -  \frac{\bar U \left(\eth (J \bar J) - 2  J \eth \bar J\right)}{8(1 + K)}  - \frac{\bar J  \partial_u J}{4 (1 + K)} + \frac{J\partial_u \bar J}{4 (1 + K)} + \frac{1}{4} \bar U \Theta - \frac{1}{4} U \bar \Theta\bigg].
  \end{align}
\end{subequations}
}
\end{widetext}

Note that the Newman-Penrose spin coefficients themselves involve no time
derivatives beyond those that would be computed during the Bondi coordinate
characteristic evolution, so they may be computed in a practical numerical
implementation with little trouble.

\subsection{Computation of Weyl scalars}

Having established the spin coefficients in terms of Bondi quantities, we can
easily derive the Weyl scalars in the Newman-Penrose formalism by taking
advantage of a selected subset of the Newman-Penrose equations,
\begin{subequations} \label{eq:weyl_scalar_identities}
\begin{align}
  \Psi_0 =& D \sigma  - \delta \kappa - (\rho + \bar \rho) \sigma - (3 \epsilon -\bar \epsilon) \sigma \notag\\
          &+ (\tau - \bar \pi + \bar \alpha + 3 \beta_{\text{NP}}) \kappa, \\
  \Psi_1 =& D \beta_{\text{NP}} - \delta \epsilon  - (\alpha + \pi) \sigma - (\bar \rho - \bar \epsilon) \beta_{\text{NP}} \notag\\
          &+ (\mu + \gamma) \kappa + (\bar \alpha - \bar \pi) \epsilon,\\
  \Psi_2 =& D\mu - \delta \pi - (\bar \rho - \epsilon - \bar \epsilon) \mu - \sigma \lambda \notag\\
          &+ (\bar \alpha - \beta_{\text{NP}} - \bar \pi) \pi + \nu \kappa,\\
  \Psi_3 =& D \nu -  \Delta \pi - (\pi + \bar \tau) \mu - (\bar \pi + \tau) \lambda \notag\\
          &- (\gamma - \bar \gamma) \pi + (3 \epsilon + \bar \epsilon) \nu, \\
  \Psi_4 =& -\Delta \lambda + \bar{\delta}\nu - \lambda(\mu + \bar{\mu})
            - (3 \gamma - \bar{\gamma}) \lambda\notag\\
          &+ (3 \alpha + \bar{\beta}_{\text{NP}} + \pi - \bar{\tau}) \nu,
\end{align}
\end{subequations}
where we have specialized to a vacuum solution, setting $R_{\mu \nu} = 0$.

While some of the spin coefficients depend on the dyad connection $\Theta$, the
coordinate invariance of the Weyl scalars ensures that all such contributions
must cancel or produce dependence on the gauge-invariant spherical curvature
  scalar in the calculation of the Weyl scalars.
This identity is used as a check on the spin coefficients calculation in the
accompanying Mathematica document \cite{companion_package}.

Most of the Newman-Penrose spin coefficients in (\ref{eq:spin_coefficients}) are
spin-weighted scalars in the Bondi-like system.
The exceptions are $\beta_{\text{NP}}$, $\alpha$, and $\gamma$, which have
explicit coordinate dependence via the connection terms $\Theta$ and
$\bar \Theta$.
We therefore introduce the more convenient method of defining new spin-weighted
scalars $\beta_{\text{NP}}^{\text{SW}} = \beta|_{\Theta = 0}$,
$\alpha^{\text{SW}} = \alpha|_{\Theta = 0}$, and
$\gamma^{\text{SW}} = \gamma|_{\Theta = 0}$.
Then, define the spin-weighted generalizations of the scalar derivatives:
\begin{subequations}
\begin{align}
  \Delta^{\text{SW}} &= \sqrt{2} e^{-2 \beta} \left(\left(\delta^\mu{}_u - \frac{V}{2 r} \delta^\mu{}_r\right)\nabla_\mu  + \frac{1}{2} \bar U \eth + \frac{1}{2} U \bar \eth\right)\\
  \delta^{\text{SW}} &= - \frac{1}{\sqrt{2} r} \left(\sqrt{\frac{K + 1}{2}}\eth - \frac{1}{2(1 + K)} J \bar\eth\right).
\end{align}
\end{subequations}
The Weyl scalar identities~(\ref{eq:weyl_scalar_identities}) remain unchanged
under the replacements
\begin{equation}
  \{\beta_{\text{NP}}, \alpha, \gamma, \Delta, \delta\}\rightarrow \{\beta_{\text{NP}}^{\text{SW}},
\alpha^{\text{SW}}, \gamma^{\text{SW}}, \Delta^{\text{SW}},
  \delta^{\text{SW}}\}.
\end{equation}
Therefore, the most direct route to calculating $\Psi_2$, $\Psi_3$, and $\Psi_4$
in the bulk of the spacetime is to perform the above replacements of the spin
coefficients and derivatives, obtaining
{\allowdisplaybreaks
\begin{widetext}
\begin{subequations} \label{eq:PartialExpandPsi24}
  \begin{align}
    \Psi_2 &= \partial_r \mu + \frac{1}{2 r} \left(\frac{J \bar \eth \pi}{\sqrt{1+K}} - \sqrt{1 + K} \eth \pi\right) + (\bar \epsilon + \epsilon - \bar \rho )\mu  + (\bar \alpha^{\text{SW}} - \bar \pi  - \beta_{\text{NP}}^{\text{SW}}  ) \pi - \lambda \sigma,\\
    \Psi_3 &= \frac{1}{\sqrt{2}} \partial_r \nu - \sqrt{2} e^{-2 \beta} \partial_u \pi +  \frac{e^{-2 \beta}}{\sqrt 2}(r W + 1) \left( \partial_r \pi - U \bar \eth \pi - \bar U \eth \pi\right)  \notag\\
            &\hspace{.5cm}- (\pi + \bar \tau) \mu - (\bar \pi + \tau) \lambda - (\gamma^{\text{SW}} - \bar{\gamma}^{\text{SW}}) \pi + (3 \epsilon + \bar{\epsilon}) \nu,\\
    \Psi_4 &= \frac{1}{2 r} \left(\sqrt{1 + K}\bar \eth \nu - \frac{\bar J \eth \nu}{\sqrt{1 + k}}\right) - e^{-2 \beta} \partial_u \lambda + \frac{e^{-2 \beta}}{2}(r W + 1) \left( \partial_r \lambda- U \bar \eth \lambda - \bar U \eth \lambda\right)\notag\\
&\hspace{.5cm}    - (\mu + \bar \mu + 3 \gamma^{\text{SW}} - \bar \gamma^{\text{SW}}) \lambda+ (3 \alpha^{\text{SW}} + \bar \beta_{\text{NP}}^{\text{SW}} + \pi - \bar \tau)\nu
  \end{align}
\end{subequations}
These can be easily expanded using the previous definitions, and the full
expression in terms of spin-weighted quantities is given in the accompanying
Mathematica document \cite{companion_package}.
However, the results do not give interesting simplifications.
The quantities $\Psi_0$ and $\Psi_1$, however, possess cancellations that allow
for a concise full expression in terms of Bondi quantities
\begin{subequations} \label{eq:FullExpandPsi01}
  \begin{align}
    \Psi_0 &= \left(\frac{r \partial_r \beta - 1}{4 K r}\right)\left((1 + K) \partial_r J - \frac{J^2 \partial_r \bar J}{(1 + K)}\right) + \frac{J (1 + K^2) \partial_r J \partial_r \bar J}{8 K^3}\notag\\
            &\hspace{.5cm} + \frac{1}{8 K}\left(\frac{J^2 \partial_r^2 \bar J}{1 + K} - (1 + K) \partial_r^2 J \right) + \frac{-J \bar J^2 \partial_r J^2 - J^3 \partial_r \bar J^2}{16 K^3},\\
    \Psi_1 &= \frac{1}{4 \sqrt{2(K+1)} r} \left(- J \partial_r(2 \bar \eth \beta - \bar Q)  + \bigg(\frac{J^2 \partial_r \bar J - (1 + K)  \partial_r J}{4 K} + \frac{J}{r}\right) (2 \bar \eth \beta  + \bar Q)\notag\\
    &\hspace{3cm}-(1+K)\partial_r \left(2 \eth \beta  - Q\right)-\left(\frac{1 + K}{r} + \frac{(1 + K)\bar J \partial_r J}{4 K} - \frac{J^2 \bar J \partial_r \bar{J}}{4K(1+K)}\right)\left(2 \eth \beta + Q\right)\bigg).
  \end{align}
\end{subequations}
\end{widetext}}
\subsection{Asymptotic Weyl scalars in Partially Flat gauge}

The full set of Weyl scalars evaluated at $\mathcal{I}^+$ would provide
detailed, gauge-invariant (up to tetrad ambiguity) information about the
dynamical spacetime.
It is therefore valuable to describe the computation of the leading asymptotic
contribution for each Weyl scalar in the partially flat gauge presented in
this paper, as well as the equivalent expressions in a true Bondi-Sachs gauge.
According to the peeling theorem, the radial falloff of the Weyl scalars obeys
\begin{equation}
  \Psi_n \sim r^{n-5}.
\end{equation}

In a general gauge, the form of the Weyl scalars cannot be significantly
simplified beyond an asymptotic expansion of (\ref{eq:PartialExpandPsi24}) and
(\ref{eq:FullExpandPsi01}) in powers of $r^{-1}$.
However, in the partially flat gauge (see Section
\ref{sec:bondi_transforms}), substantial simplifications are available.
Applying the partially flat gauge conditions and components of the
Einstein field equations, we find:
{\allowdisplaybreaks
\begin{widetext}
\begin{subequations} \label{eq:if_weyl_scalars}
  \begin{align}
    \lim_{\hat{r} \rightarrow \infty} \hat{r}^5 \Psi_0^{\text{PF}}
    &= \frac{3}{2}\left(\frac{1}{4}\ifc{\bar J}^{(1)} \ifc J^{(1)} {}^2 -\ifc J^{(3)}\right),\\
    \lim_{\hat{r} \rightarrow \infty} \hat{r}^4 \Psi_1^{\text{PF}}
    &= \frac{1}{8}\left(-12  \ifc \eth \ifc \beta^{(2)}
      +\ifc J^{(1)} \ifc{\bar{Q}}^{(1)} + 2\ifc Q^{(2)}\right),\\
    \lim_{\hat{r} \rightarrow \infty} \hat{r}^3 \Psi_2^{\text{PF}}
    &= -\frac{e^{-2 \ifc \beta^{(0)}}}{4}\left(e^{2 \ifc\beta^{(0)}} \eth \ifc{\bar Q}^{(1)} + \ifc \eth \ifc{\bar U}^{(2)} + \ifc{\bar \eth} \ifc U^{(2)} +\ifc J^{(1)} \ifc{\bar \eth} \ifc{ \bar U}^{(1)} + \ifc J^{(1)} \ifc{\bar H}^{(1)} -2 \ifc W^{(2)}\right), \\
    \lim_{\hat{r} \rightarrow \infty} \hat{r}^2 \Psi_3^{\text{PF}}
    &= 2 \ifc{\bar \eth} \ifc \beta^{(0)} + 4 \ifc{\bar \eth} \ifc \beta^{(0)} \ifc{\bar \eth} \ifc \eth \ifc \beta^{(0)} + \ifc{\bar \eth} \ifc{ \bar \eth} \ifc \eth \ifc \beta^{(0)} + \frac{e^{-2 \ifc \beta^{(0)}}}{2} \ifc \eth \ifc{\bar H}^{(1)} - e^{-2 \ifc \beta^{(0)}} \ifc \eth \ifc \beta^{(0)} \ifc \bar H^{(1)},\\\
    \lim_{\hat{r} \rightarrow \infty} \hat{r} \Psi_4^{\text{PF}}
    &= -e^{-2 \ifc \beta^{(0)}} \partial_{\ifc u} \left[
      e^{-2 \ifc \beta^{(0)}} \left(\ifc{\bar{\eth}}\ifc{\bar{U}}^{(1)}
      + \ifc{\bar{H}}^{(1)}\right)\right].
  \end{align}
\end{subequations}
In a full Bondi-Sachs gauge, the leading Weyl scalars can be simplified somewhat
further, giving
\begin{subequations} \label{eq:bondi_weyl_scalars}
  \begin{align}
    \lim_{\bs r \rightarrow \infty} \bs r^5 \Psi_0^{\text{Bondi}}
    =&  \frac{3}{2}\left(\frac{1}{4}\bs{\bar J}^{(1)} \bs J^{(1)} {}^2 -\bs J^{(3)}\right)\\
    =&  \Psi_0^{\rm PF(5)}(\ifc u(\bs u))
      + 2 \ifc \eth \bs u \Psi_1^{\rm PF(4)}(\ifc u(\bs u))
      + \frac{3}{4} \left(\ifc \eth \bs u\right)^2   \Psi_2^{\rm PF(3)}(\ifc u(\bs u))\notag\\
    & + \frac{1}{2} \left(\ifc \eth \bs u\right)^3  \Psi_3^{\rm PF(2)}(\ifc u(\bs u))
      + \frac{1}{16} \left(\ifc \eth \bs u\right)^4 \Psi_4^{\rm PF(1)}(\ifc u(\bs u)), \\
    \lim_{\bs r \rightarrow \infty} \bs r^4 \Psi_1^{\text{Bondi}}
    =& \frac{1}{8}\left(-12 \bs \eth \bs \beta^{(2)}
      +\bs J^{(1)} \bs{\bar{Q}}^{(1)} + 2\bs Q^{(2)}\right)\\
    =&  \Psi_1^{\rm PF(4)}(\ifc u(\bs u))
      + \frac{3}{2} \ifc \eth \bs u  \Psi_2^{\rm PF(3)}(\ifc u(\bs u))
      + \frac{3}{4} \left(\ifc \eth \bs u\right)^2  \Psi_3^{\rm PF(3)}(\ifc u(\bs u))
      + \frac{1}{8} \left(\ifc \eth \bs u\right)^3 \Psi_4^{\rm PF(1)}(\ifc u(\bs u)), \\
    \lim_{\bs r \rightarrow \infty} \bs r^3 \Psi_2^{\text{Bondi}}
    =& \frac{\bs W^{(2)}}{2} - \frac{1}{8} \bs \eth \bs \eth \bs{\bar J}^{(1)} + \frac{1}{8}\bs{\bar \eth} \bs{ \bar \eth} \bs J^{(1)} -\frac{1}{4} \bs J^{(1)} \bs{\bar H}^{(1)}\\
    =&  \Psi_2^{\rm PF(3)}(\ifc u(\bs u))
      + \ifc \eth \bs u  \Psi_3^{\rm PF(2)}(\ifc u(\bs u))
      + \frac{1}{4} \left(\ifc \eth \bs u\right)^2 \Psi_4^{\rm PF(1)}(\ifc u(\bs u)), \\
    \lim_{\bs r \rightarrow \infty} \bs r^2 \Psi_3^{\text{Bondi}}
    =& \frac{1}{2} \bs \eth \bs{\bar H}^{(1)}\\
      =& \Psi_3^{\rm PF (2)}(\ifc u(\bs u)) + \frac{1}{2}\ifc \eth \bs u^{(0)} \Psi_4^{\rm PF (1)}(\ifc u(\bs u)), \\
    \lim_{\bs r \rightarrow \infty} \bs r \Psi_4^{\text{Bondi}}
    =& -\partial_{\bs u} \bs{\bar{H}}^{(1)}\\
     =& \Psi_4^{\rm PF (1)}(\ifc u(\bs u)).
  \end{align}
\end{subequations}
\end{widetext}}
In the second line of each of the above identities, we have included the
relationship to the Weyl scalars computed in the partially flat coordinates.
These relations are the transformations given in \cite{Boyle:2015nqa},
adapted to our conventions for $\eth$ and Newman-Penrose tetrads.
 
\section{Conclusions} \label{sec:conclusions}

In this paper, we have laid the formulation groundwork for the next generation
of Cauchy-characteristic evolution code.
Spectral methods have shown that they give an efficient treatment of a fully
nonlinear asymptotic characteristic system \cite{Barkett:2019uae} in general
relativity.
However, they have previously suffered from logarithmic
dependence that threatens the numerical precision of polynomial-based spectral
techniques.
We have streamlined the previously derived \cite{Bishop:1997ik, Handmer:2014qha}
characteristic evolution equations for ease and efficiency of numerical
implementation (Section \ref{sec:compactified_evolution}).
We have demonstrated that there exists a computationally simple ``partially
flat'' gauge (Section \ref{sec:bondi_transforms}) that provably avoids any
logarithmic dependence (Section \ref{sec:regularity_preservation}).
We have also provided a direct roadmap (Section \ref{sec:computation_roadmap})
for the numerical implementation of our suggested method.

In the process of determining the computationally preferred partially flat
gauge, we have also extended the coordinate methods for interfacing between a
generic coordinate system and the highly fixed Bondi-Sachs coordinates.
When taken together with previous results
\cite{Bishop:1998uk} that determine a Bondi-like metric
(\ref{eq:BondiLikeMetric}) from an arbitrary coordinate system, our coordinate
transformations in Section \ref{sec:bondi_transforms}
determine the corresponding Bondi-Sachs metric, which is unique up to residual
BMS freedom.

Finally, we have demonstrated the significant simplifications given by using the
partially flat gauge for the determination of asymptotic quantities.
We provide simple formulas for computing the leading contributions to each of
the Weyl scalars (Section \ref{sec:weyl_scalars}) and the Bondi news (Section
\ref{sec:news}).

The most direct extension of this work is the numerical implementation itself,
which is underway as a module of the SpECTRE \cite{Kidder:2016hev} simulation
codebase.
In forthcoming work, we will present the performance and precision tests of the
implementation.

The ability to determine asymptotic quantities to high numerical precision
without gauge ambiguities will have numerous applications.
It can be used to develop improved waveform models for LIGO data analysis.
It can also be used for investigating various theoretical issues in general
relativity involving quantities at $\mathcal{I}^+$, including memory effects.

In future work, we will also explore the use of the methods presented in this
paper in a Cauchy-characteristic matching (CCM) scheme.
CCM uses the nonlinear characteristic evolution to provide a fully physical
vacuum boundary condition for the Cauchy code.
CCM, however, involves the considerable complexity of running an efficient
characteristic evolution in tandem, and in communication with, a Cauchy code
\cite{Szilagyi:2000xu, Bishop:1998uk}.
Because of the technical challenges, there have been no implementations of
nonlinear CCM, but the formulation advances presented in this work will ease the
way to a successful matching system.

\section*{Acknowledgments}
We thank Kevin Barkett, Michael Boyle, Dante Iozzo, Bela Szilagyi, and Jeff Winicour for valuable
discussions and suggestions regarding this project.
We thank Harald Pfeiffer for suggestions improving the presentation in this paper.
We thank Sizheng Ma for his attention to detail in
pointing out equation errors in the first published version of this work.
This work was supported in part by the Sherman Fairchild Foundation
and by NSF Grants No. PHY-1708212, No. PHY-1708213, and No. OAC-1931266
at Caltech and
NSF Grants No. PHY-1912081 and No. OAC-1931280 at Cornell.

\appendix

\section{Coordinate system glossary} \label{app:indices}

\begin{table*}[t] \label{tab:indices}
  \def\arraystretch{1.3}
\vspace{5mm}
\begin{tabularx}{\textwidth}{|Y|c|c|c|Y|}
  \bottomrule
\textbf{Coordinate system} & \textbf{Coordinates} & \textbf{Index style} & \textbf{Sections used} & \textbf{Explanation} \\
  \toprule
  \bottomrule
  Cauchy coordinates &$\{t^\prime, r^\prime, x^\prime{}^{A^\prime}\}$& $\alpha^\prime$&\ref{sec:foundations}& Input coordinates on the worldtube, provided by a Cauchy simulation.\\\hline
  Radial null coordinates &$\{\rn{u}, \rn{\lambda}, \rn{x}^{\rn{A}}\}$& $\rn{\alpha}$& \ref{sec:boundary_transforms}& Intermediate step in deriving Bondi-like coordinates. Satisfies $g_{\rn{\lambda} \rn{\lambda}} = g_{\rn{\lambda} \rn{A}} = 0$. \\\hline
  Bondi-like coordinates &$\{u, r, x^{A}\}$& $\alpha$& \ref{sec:foundations}-\ref{sec:weyl_scalars}& Coordinates in evolution of standard CCE algorithm. Satisfies $g_{r r} = g_{r A} = 0$ and $\det(g_{A B}) = \det(q_{A B})$\\ \hline
Partially flat \newline Bondi-like coordinates & $\{\ifc{u}, \ifc{r}, \ifc{x}^{\ifc{A}}\}$
  & $\ifc{\alpha}$& \ref{sec:bondi_transforms}, \ref{sec:regularity_preservation}, \ref{sec:weyl_scalars}& Partially restricted Bondi-like coordinates, preferred for use in computational schemes described in this paper. Satisfies $g_{\ifc r \ifc r} = g_{\ifc r \ifc A} =0$, $\det(g_{\ifc A \ifc B}) = \det(q_{\ifc A \ifc B})$, and all metric components except for $\ifc\beta$ asymptotically approach Minkowski form.\\ \hline
  Bondi coordinates & $\{\bs u, \bs r, \bs x^{\bs A} \}$& $\bs \alpha$&\ref{sec:foundations}, \ref{sec:bondi_transforms}, \ref{sec:weyl_scalars}&  Bondi-Sachs coordinates, satisfying $g_{\bs r \bs r} = g_{\bs r \bs A} = 0$, $\det g_{\bs A \bs B} = \det q_{\bs A \bs B}$, and asymptotically approaches Minkowski as per (\ref{eq:BondiSachsFalloffs}).\\\hline
  Numerically adapted \newline coordinates & $\{\na u, \na y, \na x^{\na A}\}$ & $\na{\alpha}$&\ref{sec:compactified_evolution}& Coordinates associated with a generic Bondi or Bondi-like coordinate system (which can include the partially flat or true Bondi-Sachs coordinates), but with a numerically adapted radial coordinate $\na y \in [-1, 1]$.\\\toprule
\end{tabularx}
\caption{Coordinate and index notation used in our presentation of the CCE formalism.}
\end{table*}

Because of the ease of generating subtle mistakes or misunderstandings when
using multiple similar coordinate systems, we are careful throughout this paper
to introduce a complete coordinate system with distinct index style each time a
coordinate transformation is suggested.
In scenarios involving partial derivatives, the coordinate notation carries the
implication of holding fixed all remaining coordinates of a given set.
The need to clearly express derivatives motivates our extensive notation.
For instance, the Cauchy coordinates and Bondi-like coordinates share time and
angular coordinates $\bl{u} = t^\prime$,
$\bl{x}^{\bl{A}} = \delta^{\bl{A}}{}_{A^\prime} x^\prime{}^{A^\prime}$, but
because of the differing radial coordinates, the partial derivatives with
respect to time and angles are not the same, e.
g.
$g_{\bl{A} \bl{B}, \bl{u}} \ne g_{A^\prime B^\prime, t^\prime}$.
The full collection of our coordinate system notation is given in Table \ref{tab:indices}.

\section{Additional transformations to  partially flat coordinates} \label{sec:IF_extras}

Here we give detailed transformations for $\ifc Q$ and $\ifc H$ in a form
amenable to numerical implementation for boundary computations described in
Section \ref{sec:regularity_preservation}.

$\ifc Q$ may be obtained from $\partial_{\ifc r} \ifc U$ as:
\begin{equation} \label{eq:ifcQ}
\ifc{Q} = \ifc r^2 e^{-2 \ifc \beta} (\ifc K \partial_{\ifc r} \ifc U + \ifc J \partial_{\ifc r} \ifc{\bar U}),
\end{equation}
and
\begin{widetext}
\begin{align}
  \partial_{\ifc r} \ifc U =& \frac{1}{2 \ifc \omega^3}\left(\ifc{\bar b} \partial_r U - \ifc a \partial_r \bar U\right) + \frac{e^{2\ifc \beta}}{4 \ifc \omega} \left(\ifc J \ifc{\bar{\eth}} \ifc \omega - \ifc K \ifc \eth \ifc \omega\right)\left(\partial_{\ifc r} \ifc{\bar{J}} \partial_{\ifc r} \ifc J - \frac{\partial_{\ifc r}(\ifc J \ifc{\bar{J}})^2}{4 \ifc K^2}\right)\notag\\
  &+ \frac{e^{2 \ifc \beta}}{\ifc \omega \ifc r} \left[ \ifc{\bar{\eth}} \ifc \omega \left(\partial_{\ifc r} \ifc J-\frac{\ifc J}{\ifc r}\right) + \ifc{\bar{\eth}} \ifc \omega \left(\frac{\ifc K}{\ifc r} - \frac{\ifc J \partial_{\ifc r} \ifc{\bar J} + \ifc{\bar J} \partial_{\ifc r} \ifc J}{2 \ifc K}\right) \right].
\end{align}
In terms of other transformed quantities, $\ifc H$ is
\begin{align} \label{eq:ifcH}
  \ifc H =& \frac{\partial_{\ifc u} \ifc \omega - \tfrac{1}{2} \left(\mathcal{U}^{(0)} \bar{\ifc \eth}\ifc \omega + \bar{\mathcal{U}}^{(0)} \ifc \eth \ifc \omega \right) }{\ifc \omega} \left(2 \ifc J - \ifc r \partial_{\ifc r} \ifc J\right) - \ifc J\ifc{\bar \eth} \mathcal{U}^{(0)} + \ifc K \ifc \eth \bar{\mathcal{U}}^{(0)}  \notag\\
  &+ \frac{1}{4 \ifc \omega} \left(\ifc{\bar b}^2 H + \ifc a^2 \bar H + \ifc{\bar b} a \frac{H \bar J + J \bar H}{K}\right) + \frac{1}{2} \left(\mathcal{U}^{(0)} \ifc{\bar \eth} \ifc J + \ifc{\eth}(\bar{\mathcal{U}}^{(0)} \ifc J) - \ifc J \ifc \eth \bar{\mathcal{U}}^{(0)}\right)
\end{align}
\end{widetext}

\section{Perturbative expansion of the transformations to Bondi-Sachs coordinates near $\mathcal{I}^+$} \label{app:perturbative_transformations}

Here we give the first few orders of the asymptotic coordinate transformations
necessary to move from partially flat coordinates to Bondi-Sachs coordinates.
These expressions are expanded from the equations derived in Section
\ref{sec:bondi_transforms}.

Perturbative expansion of the defining equation for
$\bs u = \bs u^{(0)} + \ifc l \bs u^{(1)} + \ifc l^2 \bs u^{(2)} +
\mathcal{O}(\ifc l^3)$ in (\ref{eq:uRequation}) gives rise to the equations
\begin{widetext}
\begin{subequations}
\begin{align}
  2  \mathring u^{(1)} &= -\eth \mathring u^{(0)} \bar \eth \mathring u^{(0)},\\
  -4 \mathring u^{(2)} - 2 e^{- 2 \hat \beta} \partial_{\hat u} \mathring u^{(1)} \mathring u^{(1)} &= 2 \partial_{\hat A} \mathring u^{(0)} \mathring u^{(1)} \hat U^{(1) \hat A} e^{-2 \hat \beta} - \left(\bar \eth \mathring u^{(0)}\right)^2 J^{(1)} - \left(\eth \mathring u^{(0)}\right)^2 \bar J^{(1)}.
\end{align}
\end{subequations}

Perturbative expansion of
$\bs x^{\bs A} = \delta^{\bs A}{}_{\ifc A} \ifc x^{\ifc A} + \ifc l \bs x^{(1)
  \bs A} +\ifc l^2 \bs x^{(2) \bs A} + \mathcal{O}(\ifc l^3) $ in
(\ref{eq:bsxA}) gives the equations
\begin{subequations}
  \begin{align}
    -\bs x^{(1) \bs A} =& \ifc q^{\ifc A \ifc B} \partial_{\ifc B} \bs u^{(0)},\\
    -\bs x^{(2) \bs A} =& \partial_{\ifc u} \bs u^{(1)} e^{-2 \ifc \beta} \bs x^{(1) \bs A} + \bs u^{(1)} \partial_{\ifc u} \bs x^{(1) \bs A} e^{-2 \ifc \beta} + \delta^{\bs A}{}_{\ifc A} \ifc U^{(1) \ifc A} \bs u^{(1)} e^{-2 \ifc \beta} + \bs x^{(1) \bs A} \partial_{\ifc A} \bs u^{(0)} \ifc U^{(1) \ifc A} e^{-2 \ifc \beta}\notag\\& + \partial_{\ifc B} \bs u^{(1)} \delta_{\ifc A}{}^{\bs A} \ifc q^{\ifc A \ifc B} + \partial_{\ifc B} \bs u^{(0)} \partial_{\ifc A} \bs x^{(1) \bs A} q^{\ifc A \ifc B} + \partial_{\ifc B} \bs u^{(0)} \delta{}_{\ifc A}{}^{\bs A} h^{(1) \ifc A \ifc B}.
  \end{align}
\end{subequations}
\end{widetext}
Finally, the leading alteration to the conformal factor  $\bs \omega = 1 + \ifc l \bs \omega^{(1)} + \mathcal{O}(\ifc l^2)$ is
\begin{subequations}
  \begin{align}
\bs    \omega^{(1)} = \frac{1}{2} \ifc \eth \ifc{\bar \eth} \bs u^{(0)}.
  \end{align}
\end{subequations}

\bibliography{characteristic_formulation}

\bibliographystyle{apsrev4-2}

\end{document}